\documentclass[journal]{IEEEtran}

\usepackage{xcolor,soul,framed} 

\colorlet{shadecolor}{yellow}
\usepackage[pdftex]{graphicx}
\DeclareGraphicsExtensions{.pdf,.jpeg,.png}

\usepackage[cmex10]{amsmath}
\usepackage{textcomp}
\usepackage{array}
\usepackage{mdwmath}
\usepackage{mdwtab}
\usepackage{eqparbox}
\usepackage{url}
\usepackage{comment}
\usepackage{subcaption}
\usepackage{float}
\usepackage{tabularx}
\usepackage{makecell}
\usepackage{algorithm}
\usepackage{algpseudocode}

\begin{document}
\bstctlcite{IEEEexample:BSTcontrol}

\title{Parallelobox: Improved Decomposition for Optimized Parallel Printing using Axis-Aligned Bounding Boxes}
\author{Hayley Hatton,
      Muhammed Khalid,
      Umar Manzoor,
      John Murray
\thanks{
        H. Hatton \& M. Khalid are with the School of Digital \& Physical Sciences, University of Hull, Hull, HU6 7RX, UK. U. Manzoor is with University of Wolverhampton, UK. J. Murray is with University of Huddersfield, UK. \\
        
  Correspondence: email: m.khalid@hull.ac.uk
}
}


\markboth{
}{}

\maketitle

\begin{figure*}[ht]
    \frame{\includegraphics[width=\textwidth]{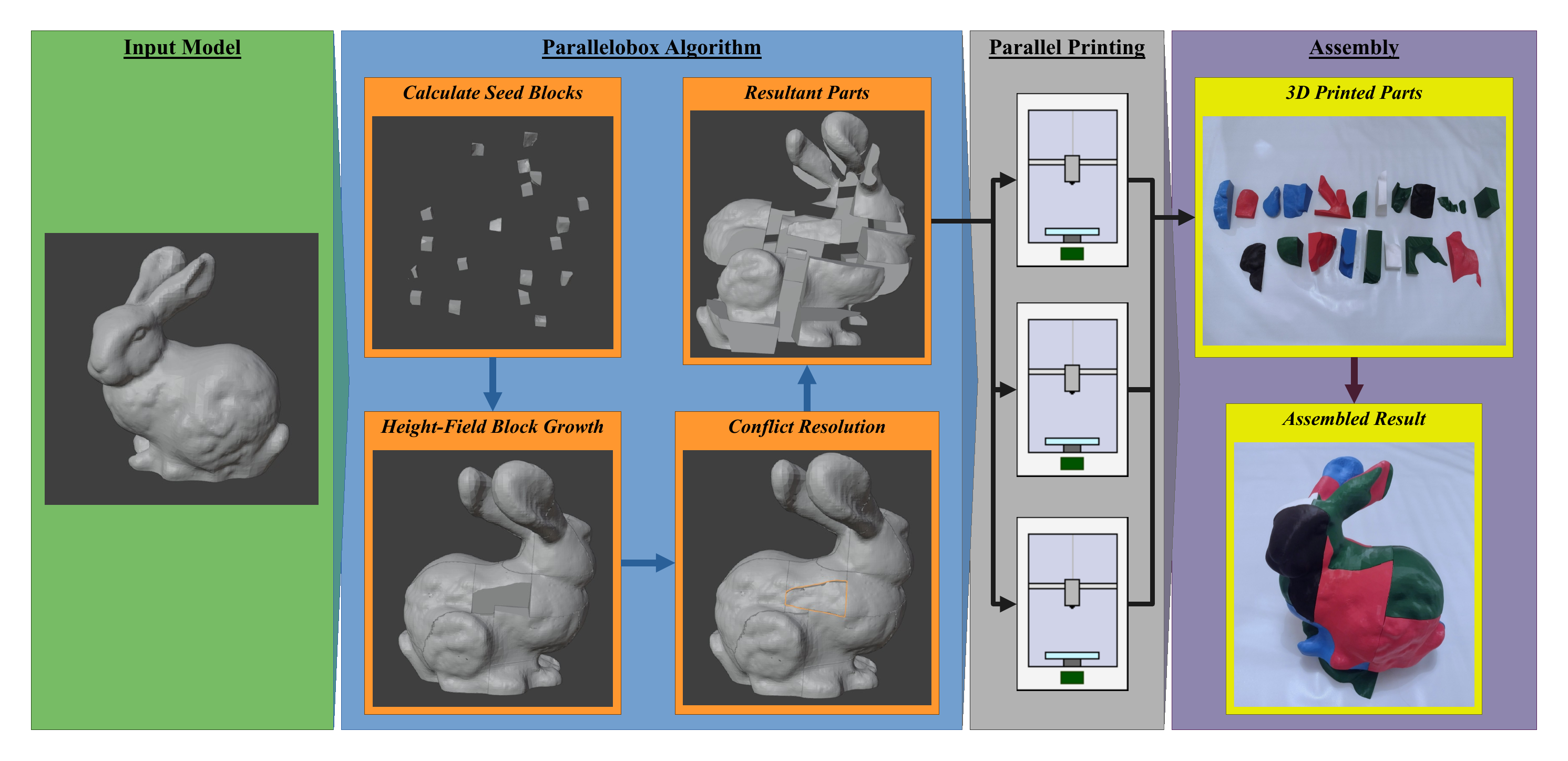}}
    \captionof{figure}{Overview of the \textit{Parallelobox} process, from model to printing and assembly.}
    \label{fig:overview}
\end{figure*}

\begin{abstract}
Much contemporary research in additive manufacturing focuses on breaking down models into constituent parts in the pursuit of various factors, such as printability of large models in smaller printing volumes, or reduction of support structures. Newer research has begun to focus on using these decomposition processes for printing models across multiple printers in parallel. We present a novel approach to this that incorporates axis-aligned bounding boxes as height fields to improve the characteristics of decomposition, including printing time, feasibility, and aesthetics. By expanding these bounding boxes according to a parallel printing objective, with additional improved efficiency from a metaheuristic process, these boxes can then be used for rapid decomposition using simple out-of-the-box mesh clipping operations. This algorithm is experimentally evaluated across a range of models against two other contemporary approaches to parallel printing that use more rudimentary techniques, such as recursive symmetry and cube skeletonization. \textit{Parallelobox} outperformed each of these across a range of sample models on the basis of a "parallel printing time" metric using simulated 3D printing to compute the results.
\end{abstract}


%
\IEEEpeerreviewmaketitle


\section{Introduction}

\IEEEPARstart{I}{n} the field of additive manufacturing, "decomposition" (also known as "partitioning") is often used to break apart a model into smaller pieces for various purposes, such as, for example, enabling the model to be printed in a smaller printing volume than the full size of the model, or reducing the support structures required to print the model. 

A more novel application of decomposition that has emerged in recent years involves utilizing multiple sets of 3D printers to print parts of models simultaneously in parallel to reduce the real time required to print them \cite{Chen:2016} \cite{Lamboo:2021}. This application imposes different constraints and requirements on the decomposition process. Where other algorithms may freely tolerate the production of decomposed parts with large disparities in size and surface area (should this be acceptable in pursuit of their own objectives), and may instead focus on reducing the aggregate printing time, in the case of parallel printing, printer utilization is most important. What this means is that a longer aggregate printing time is perfectly acceptable if it is translated into a shorter parallel printing time \cite{Hatton:2023}.

In an ideal algorithm, given any number of printers, a perfectly even distribution of printing time across the decomposed parts can be rapidly computed without any other side-effects (for example, impairments in the final assembly's aesthetic). This would effectively mean that all available printers are utilized perfectly, with no redundant printer time whatsoever. However, there are many different forms this can take, as there are potentially infinite ways to decompose models subject to this singular constraint. This opens up the potential for approaches that can achieve this as best as possible given certain desired parameters, while potentially improving other aspects of the process as well.

In this paper, an alternative approach to decomposition is proposed. The concept of "axis-aligned bounding boxes" (AABBs) is utilized as a core tool for guiding the partitioning process. These axis-aligned bounding boxes can effectively become quadratic "height field" blocks that represent the surface of a model when arranged properly across it. When represented this way, height field blocks supply desirable properties to the algorithm, such as reduced overhang and the possibility of eliminating significant quantities of internal volume. This can primarily reduce printing time and material usage while also improving aesthetic quality. Their axis-aligned nature constrains the complexity of their implementation and makes their application considerably more feasible than oriented boxes, which are much more complicated to arrange. 

This paper presents a novel algorithm that combines the distribution of axis-aligned bounding boxes across input models with an objective that produces a decomposition set with minimal parallel printing time, while secondarily reaping the additional benefits from using AABBs in aesthetic quality. This algorithm is compared to two other contemporary papers in the field of geometry decomposition for parallel printing, using simulated data across a disparate range of valid models. The results demonstrate the temporal superiority of this algorithm in comparison, where this algorithm was considerably more efficient in general and more flexible in how it tolerates a range of different geometries. This is achieved through the efficient positioning and expansion of height blocks that can be aware of the surface properties of the geometry, as compared to the more top-down approaches of the other two papers.

This paper offers the following technical contributions:

\begin{itemize}
    \item An algorithm for decomposing models for printing simultaneously across a given number of printers that, at its core, uses axis-aligned blocks as height fields for a flexible and even distribution of mass while reducing overhanging areas.
    \item Further refinements to the performance of the algorithm using a symmetry-based decomposition step, \textit{k-means++} clustering for initial block placement, and a metaheuristic for printer quantity determination.
    \item An experiment comparing this algorithm with two pre-existing "parallel partitioning" algorithms, demonstrating the superiority of this approach in reducing the "parallel printing time".
\end{itemize}

\section{Related Work}

\subsubsection{Decomposition for Additive Manufacturing}

This paper branches from the higher domain of research that involves the optimization of additive manufacturing processes, specifically the subdomain that involves \textit{decomposition} or \textit{partitioning} models into constituent parts. One of the defining papers for this is \textit{Chopper} \cite{Luo:2012}, which approaches the familiar problem of having to fit models into smaller printing volumes than themselves by decomposing them into smaller parts. To this end, it employs binary space partitioning (BSP) strategies to decompose models through planar cuts. The substance of the algorithm is in arranging these BSPs in a way that best fits the objective of printability, while optimizing other qualities such as structural stability, connection strength, and aesthetics (e.g. hiding the seam). Prioritization is achieved through the evaluation of a set of objective functions which rank the BSPs in terms of these qualities.

Another, more recent paper showing evolution from this archetype, is \textit{Optimizing Object Decomposition to Reduce Visual Artifacts in 3D Printing} \cite{Filoscia:2020}, which tackles the same issue of breaking models down into smaller parts to be printed, but iterates on the limitations of planar cuts by using "non-planar cuts" instead. This process begins with oversampling patches of the model surface, whose boundaries demarcate potential cut regions. A labeling "Integer Linear Program" is applied to refine this initial segmentation, allowing cuts to deviate in such a way as to preserve appearance from seams by directing them through regions identified as "occluded" through the use of an ambient occlusion mapping function, while also incentivizing against overhanging regions to reduce the need for supports. The quantity of cuts required is ultimately selected based on the size of the model. Their reasoning is that a greater number of cuts only makes sense for larger models, as smaller models are better printed and assembled in fewer pieces. Their results demonstrate this using both quantitative and qualitative assessments on a range of largely complex 3D models, but lacked in ensuring printability due to the lack of maximum size limitations of produced parts, and ensuring the feasibility of reassembling the pieces.

Other research has explored decomposition in many different directions. For example, \textit{Skeleton Partition Models for 3D Printing} \cite{Chen:2022} and \textit{Models Partition for 3D Printing Objects Using Skeleton} \cite{Jiang:2017} both explore making decomposition more efficient in the domains where models feature appendages, such as skeletons, by exploiting the presence of concave surface features to identify parts of an object such as limbs, and use these as points to locate cuts for decomposing an object. The intersection points between the appendages and the core is a natural area to direct cuts, particularly in the preservation of aesthetic quality. \textit{Automatic Segmentation and 3D Printing of A-shaped Manikins using a Bounding Box and Body-feature Points} \cite{Jung:2021} takes the same principle, but uses bounding boxes instead as the mode of decomposition to constrain parts to printing volumes.

\subsubsection{Decomposition in Parallel} 

Using decomposition within the context of parallel printing is a somewhat novel concept. Three papers currently represent this interest in the literature.

\textit{Harpe: Partitioning Models to Minimize Parallel Print Time} \cite{Lamboo:2021} is the first major implementation of the idea in the domain. Building off the prior \textit{Chopper} \cite{Luo:2012}, it involves the use of binary space partitioning to break down a model using planes to recursively "cut" parts into two, but unlike \textit{Chopper} performs this in a parallel-aware context. A heuristic was developed which approximates print time from volume and area, and the algorithm uses this to select the best cuts at each stage using a beam search, with the best cuts being those that reduce the maximum approximated printing time of each sub-part. "Batching" of computations was additionally implemented to improve the calculation performance. The final results were largely simulated and found improvements in parallel print time compared to Chopper, with diminishing returns with increased parallelization.

\textit{Symmetry-Based Decomposition For Optimised Parallelisation in 3D Printing Processes} \cite{Hatton:2023} (hereafter referred to as \textit{"Symmetry-Based Decomposition"} or "SBD"), on the other hand, involves the recursive application of planar slices located at planes of symmetry, or the closest thereof. In theory, symmetry would allow for optimal parallelized efficiency, with perfectly evenly distributed volumes in the best cases. This research involved a significant quantitative analysis in the real world of parallel printing performance directly compared to traditional serial printing methods, and found significant improvements in real-world printing times that heavily diminished with increased parallelization. This found symmetry as useful but inadequate, often with visual deficits and support requirements, as well as more broadly finding that intelligent decomposition was unnecessary for parallel printing on models that can be printed whole.

\textit{An Algorithm For Partitioning Objects Into A Cube Center And Segmented Shell Covers For Parallelized Additive Manufacturing} \cite{Li:2021} (hereafter referred to as \textit{"Cube Skeleton Segmented Shell"} or "CSSS") sees the novel deviation from planar slice decomposition strategies in the domain, centering on the concept of a "maximally circumscribed cube" (a cube that fits the internal volume that cannot get any bigger without violating that constraint), whose sides are extruded into surrounding cubes that encompass the model. These are selectively merged in order to improve the parallel printing time while still ensuring that they can fit a specified printing volume.

\subsubsection{Axis-Aligned Bounding Boxes}

Perhaps the largest inspiration for this project is \textit{Axis-Aligned Height-Field Block Decomposition of 3D Shapes} \cite{Muntoni:2018}. In this paper, the problems of using planar slices were mitigated by employing axis-aligned bounding boxes as height fields for decomposition instead, which allowed for better preservation of details of the external surface and the optimization of inner volumes. Initial axis-aligned bounding boxes are generated at each vertex and are iteratively grown to encompass the model. When these boxes collide, they are selectively merged in a way that causes the best reduction in support structures to improve output fidelity, while also limiting the number of overall pieces while ensuring that they can be printed in a specified printer's volume.

This project was found to produce appropriate decompositions without too much computing time and according to their constraints. Comparative data was lacking, making it an interesting proof of concept for later adoption and use.

\subsection{Contribution}

This paper extends the literature by formulating and demonstrating a novel approach to decomposition for parallel printing. By integrating the use of axis-aligned bounding boxes for decomposition with a wider parallel-aware objective, a potentially improved approach to parallel decomposition can be designed, offering the theoretical benefits of axis-aligned bounding box partitioning (such as reduced supports, improved fidelity, and safe removal of internal volume regions) but in a parallel context. The additional application of a symmetry metric should also help to further improve performance. This design will be tested against an appropriate selection of preexisting parallel decomposition algorithms.

\section{Problem Overview}

\subsection{Parallel Printing}

At the highest level, this research's objective is similar to other parallel partitioning algorithms: to attempt to optimize the method of decomposition such that its "parallel printing time" is minimized. This contrasts with the usual, non-parallel ("serial") approaches in the general literature that consider printing time that focus on improving the total printing time. This approach makes sense in that domain, where the printing times of the final model are strictly related to the aggregate of the printing times of their parts, but in this domain that is not necessarily the case. As previous research found parallelization to be most useful when applied to larger models \cite{Hatton:2023}, and related research in the area of parallelized decomposition focused on models initially larger than their printer volumes \cite{Lamboo:2021} \cite{Li:2021}, this will also be the case here.

"\textit{Parallel Printing Time}" in this context means the effective real printing time as observed by the person overseeing the prints. For example, five printers printing a part each at 20 minutes has an aggregate printing time of 100 minutes, but because the entire job is done in 20 minutes as they're all performed in parallel, the "parallel printing time" is only 20 minutes. There are three notable observations that affect parallel performance that are not relevant to the traditional serial approach:

\begin{itemize}
    \item The longest print is the bottleneck. Consider a setup where four parts take 20 minutes, but one part takes 2 hours - the full set of parts is not available for 2 hours. This results in a situation that is equivalent to the naval aphorism "the fleet moves as fast as the slowest ship". This incentivizes parallel-aware algorithms to reduce the printing cost of the most expensive part.
    \item Similarly, printers can be viewed as multi-core processors performing "work". As in that domain, the ideal situation involves 100\% printer utilization, where every printer is utilized at the exact same time. Situations where some printers are remaining idle waiting on another printer, is wasted time and inefficiency.
    \item Conversely, it does mean that the size of smaller parts is less important than the largest part with regard to the parallel printing time. Alterations to these parts could be made without degrading the performance of the decomposition, assuming that none grew larger than the largest parts.
\end{itemize}

Consequently, the algorithm's core objective is principally to reduce the printing time of the largest decomposed part, while also attempting to maintain generally efficient utilization of 3D printer work. As print time can be reasonably modeled as a function of volume and surface area \cite{Raise3D:2024}, this is approximately translated as reducing the size and surface area of this largest part and otherwise keeping the decomposed parts as equal in these variables as reasonably possible. More accurate methods exist, but these are more complex and therefore incur greater costs to computation speed and implementation effort \cite{Medina:2019}.

\subsection{Application of Axis-Aligned Bounding Boxes}

With the parallel printing domain elaborated, this leads to the more specific scope of the problem this research attempts to solve: the use of "axis-aligned bounding boxes" to achieve an optimal parallel-aware decomposition. A set of axis-aligned bounding boxes is used to decompose the model into height fields, which requires the distribution of these boxes across the surface of the model, where the surface features become the height displacements of the resulting height fields. This specifies the problem further to one of solving a geometric problem of arranging a set quantity of axis-aligned bounding boxes such that the parallel printing costs of the height field blocks are minimized; that is, the minimization of the printing cost of the largest height field block.

As a core requirement, what the algorithm needs to achieve is the production of a set of axis-aligned bounding boxes that presents a perfect 1:1 coverage of the surface geometry of the model. Lower than this, there is surface area that is not being printed, therefore the model is incompletely represented; higher than this, there is surface area that is being printed multiple times and therefore cannot be reassembled.

Compared to the paper \textit{An Algorithm For Partitioning Objects Into A Cube Center And Segmented Shell Covers For Parallelized Additive Manufacturing} \cite{Li:2021}, which can be conceptualized as a top-down approach to decomposition, this research takes more of what can be conceptualized as a bottom-up approach. An initial heavily "oversampled" regular decomposition takes place, and larger blocks are built up as composites from this set using a coarse awareness of the geometry's features. One of the main advantages of this approach is that it is more extensible and, therefore, theoretically allows for better representation of models. The former is simpler and faster, but less adaptable, so it is less able to handle more complex geometries. In contrast, this approach allows for boxes to be shaped around complex features and non-convex geometries.

\section{Method}

With the algorithm's target stated, the implementation follows to realize it. This algorithm can be envisioned on the higher level as a pipeline with multiple steps that transforms the input model into the output constituent parts for printing.

\begin{figure}
    \includegraphics[width=\columnwidth]{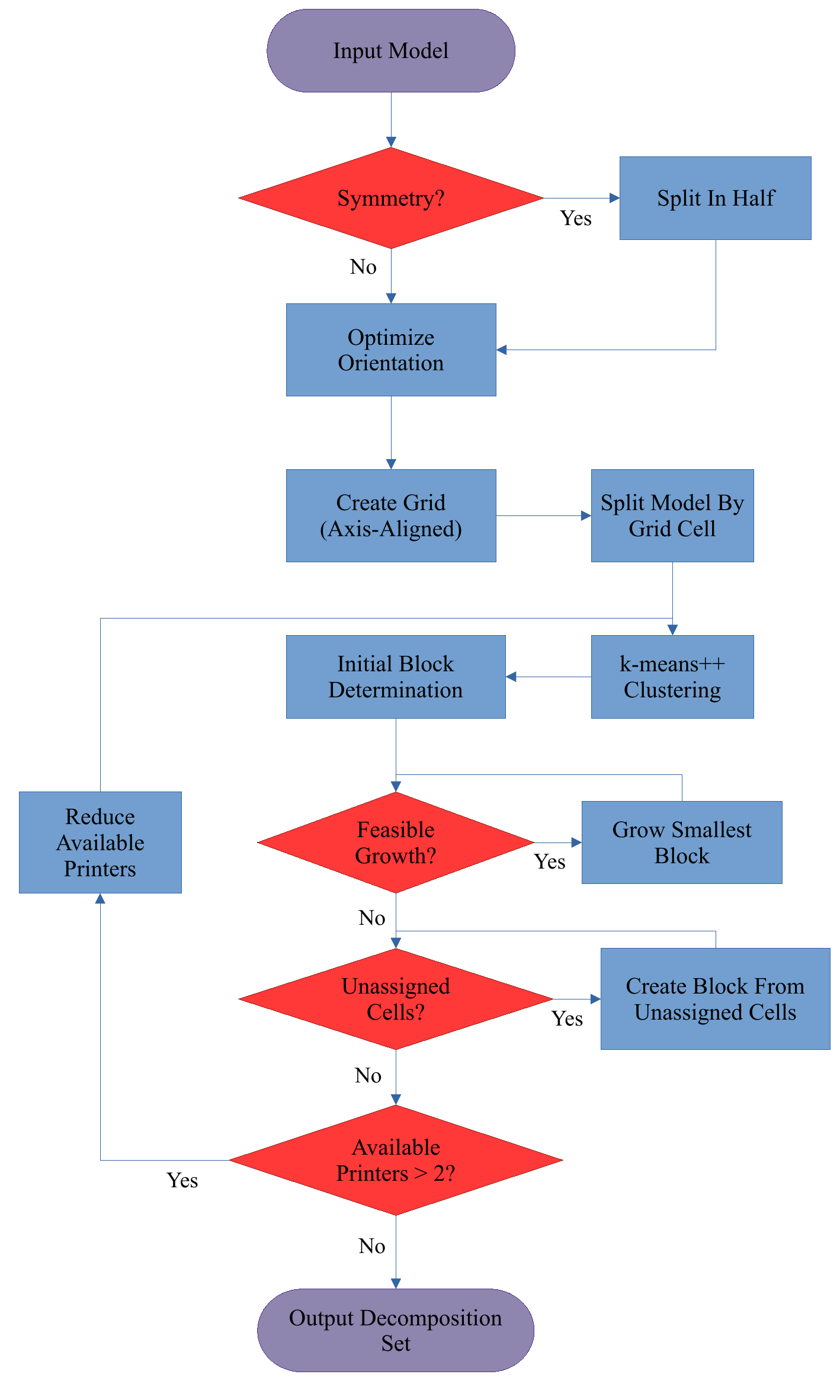}
    \captionof{figure}{Flow chart depicting the overview of the critical stages of the algorithm.}
    \label{fig:overviewflow}
\end{figure}

\subsection{Symmetry Optimization}
As discussed in previous research by Hatton et al. 2023, symmetry can offer a gold standard for parallel decomposition: as symmetry implies an even distribution of mass and surface area between the halves, the division of a model perfectly (or near perfectly) in half through its plane of symmetry represents the best possible case. Models that can be recursively subdivided in this manner would have perfectly even printing times with the amount of printers. Although previous research indicated rapidly diminishing returns, particularly as most models are not recursively symmetric, using this as an initial step can improve the performance of the algorithm.

To this end, we follow the implementation by Hatton et al. 2023 \cite{Hatton:2023} with a rudimentary symmetry determination algorithm that finds the best plane of symmetry, equipped with a score based on the error of the reflected vertex positions that determines how symmetric or asymmetric this symmetry is.

If the model is sufficiently symmetric (i.e. allowing some deviation from perfect symmetry), the planar cut is performed; if not, it is skipped, for it is not essential to the algorithm.

\subsection{Orientation Optimization}
The algorithm relies on axis-aligned bounding boxes. It follows from this that the bounding boxes are constrained in how they can be oriented across the model (this would not be the case if the algorithm used oriented bounding boxes instead). Because of this, any orientation optimization has to be done by changing the orientation of the model itself.

Different orientations can affect the decomposition of the model. As found in \cite{Vanek:2014} \cite{Wang:2020}, the orientation of models can affect the overhang requirements even if nothing else changes. To iterate from this, before applying the rest of the algorithm, it first attempts to find an optimal alignment of the model in space. This is not overly complex, reducing to a few simple goals:

\begin{itemize}
    \item Align the centroid of the model to the origin of the space. This enables an optimal distribution of the model volume through the cutting space later.
    \item Attempt to reduce overhang through orientation. Although decomposition may disturb this, it is preferential that the overhang is reduced before the process to allow the algorithm more flexibility in later steps.
    \item Attempt to align the model so that its best plane of symmetry is aligned with the fundamental planes of the slicing space. This should enable the algorithm to exploit any symmetry, as equal spaces should be distributed.
\end{itemize}

Aligning the centroid with the origin is trivial: the centroid is determined by averaging the position of all vertices and using the inverse of this position as the translation vector to apply linearly across all vertices of the model.

Orientation of the model is performed by creating an oriented bounding box of the model and determining the rotation of that based on the prior objectives. An affine transformation matrix is constructed based on the rotation from one vector to another and is applied across the model.

\subsection{Grid-Based Fine Decomposition}

From this point in the algorithm, the model is set and ready to be processed. However, in order to proceed, the initial conditions must be specified for block growth. This requires two fundamental aspects: the initial blocks and the regions into which to grow. As the algorithm relies on axis-aligned bounding boxes, and can be seen as a "bottom-up" algorithm that starts with smaller regions aggregating into larger regions (rather than taking larger regions and shrinking them down), both of these aspects can be achieved at once by using a "grid-based decomposition". 

The essence of this step is to use the cubes of each cell in the grid as the method of decomposition, through using them as boolean intersection operations on the target mesh. Boolean mesh operations are simple and fast in comparison to more complicated decomposition methods, and are available out-of-the-box in libraries such as \textit{CGAL} \cite{CGALBool:2025}, preventing effort having to be needlessly expended on an ad-hoc low-level decomposition technique where it is best spent on the broader scope. The advantages of using this approach for cutting the mesh allow decomposition operations to be more "expendable", allowing for reasonably aggressive cutting.

To this end, the grid is created by taking the maximum and minimum vertex positions of the model, scaling them slightly to enable complete coverage while avoiding rounding errors, and using these positions as the definition for a regular quad. With the regular quad established, it is transformed into a regular grid by mapping integer coordinates into model space using the base size of a cell (which is constrained to being a regular cube). The size of each cell is initially determined by dividing the volume by the input quantity of initial boxes, itself derived from a granularity parameter specified by the user, where finer settings select for higher numbers of initial boxes and therefore smaller base cell sizes. This typically results in an initial cell count between several hundred and many thousands.

Other approaches could be taken here that would not necessitate the use of a grid-based cutter, such as steadily increasing the block regions through the vertices rather than by block quanta, but this risks significant performance drawbacks in requiring more substantial computation time (particularly in larger models).

\subsubsection{Surface Region Extraction}
Once the grid-based cutter is initialized, the initial set of oversampled blocks is calculated by looping through every grid cell and cutting the mesh with it. As the original mesh itself is not modified, this step is trivial to parallelize for considerable computational improvements. The output locations are deterministic and independent, leading to a very minimal critical region that almost eliminates the synchronization time.

\begin{figure}
    \includegraphics[width=\columnwidth]{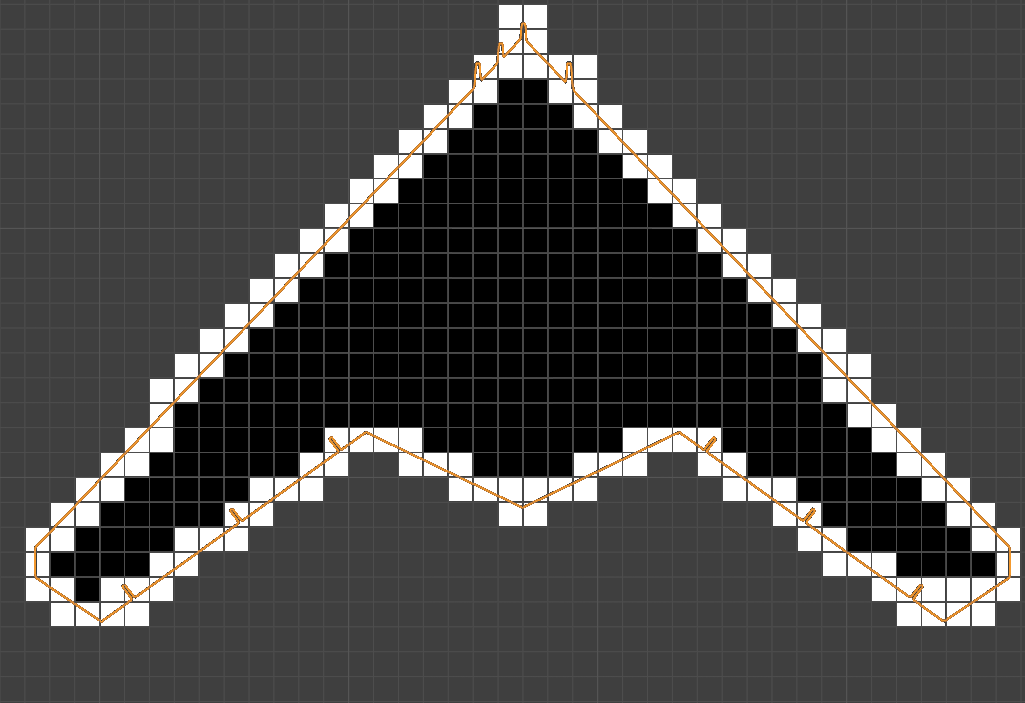}
    \captionof{figure}{2D Cross-sectional example of region classification for extraction. Grey cells are the "External" voids, Black cells are the "Internal" spaces, White cells are the "Boundary" cells.}
    \label{fig:boundaries}
\end{figure}

Subsequent steps require some domain information about the mesh in order to function, which are encoded within the initial block set. To this end, each cell is classified according to its contents relative to the model, as seen in \textit{Figure \ref{fig:boundaries}}:

\begin{itemize}
    \item \textbf{Boundary:} Cells that are on the surface of the model are marked as "boundary" cells. These cells contain surface information about their enclosed region of the model, as well as some volume information where it is internal, forming a coherent mesh themselves. They are determined by if a non-zero quantity of vertices is determined in a mesh-only cutting (without considering volume).
    \item \textbf{Internal:} Cells marked as internal are entirely contained within the volume of the model. These are determined by if there are no vertices in the mesh-only cutting, but there are vertices in the volumetric cutting.
    \item \textbf{External:} Cells marked as external do not contain any part of the model; they are the empty space surrounding it. These are determined by if there are no vertices in the boolean cut result at all.
\end{itemize}

These classifications guide the later steps. The bulk of the operations that follow take place in the boundary set, as this is the important information for producing the final model. Internal cells are avoided as much as possible, as the internal space can be hollow without affecting the final appearance of the model. External cells can be avoided completely, as growth into that region is not only useless but can result in multiple discrete regions being manipulated as a single region, which is far from desirable given the specified input printer quantity.

\subsection{Initial Block Determination}
At this stage, the model is (optionally) split in half, oriented with the base axes of 3D space, and broken down into a large set of minimal parts using a cubic lattice. The final set of decomposed parts must emerge from the algorithm as sets of these parts, which are achieved using a "block growth" strategy, as mentioned before and discussed later. However, in order to perform this, the initial "seed" blocks must be decided.

The number of seed blocks to start with is decided trivially. The quantity of decomposed parts is hard limited by the availability of printers - more decomposed parts than there are printers is a situation that is beyond the scope of the algorithm in its optimization goal. Less, however, is acceptable (and sometimes useful, as discussed later). Consequently, the number of initial seed blocks is set to the number of available printers.

Selecting the minimal parts that are seed blocks is carried out with the awareness that some configurations are likely to produce better block growth performance than others. Primarily, seed blocks should be distributed as evenly across the volume as possible, while taking the geometric complexity of areas into account, and should be relatively consistent. This makes uniform random selection of minimal parts unappetizing as an approach. Instead, \textit{k-means++} was selected, since the clustering approach combined with the heuristic should result in areas of complexity being more likely to be assigned their own blocks, while the seed blocks are distributed uniformly throughout the model. This is not entirely novel, having been employed elsewhere in the literature for similar purposes \cite{Rezaei:2018}. Alternative clustering algorithms could have been used but are typically more complex, but as this is not the core of the project \textit{k-means++} was used as it is simple and sufficient. \textit{k-means++} was selected to augment the typical \textit{k-means} as it is not much more complicated or expensive, but improves it with more spread out initial seed positions. \textit{k-means} most struggles with data sets where clusters are strongly separated, but this is not \textit{typically} the case with the vertices of models, making it appropriate for this use case. \cite{Franti:2019}

An additional constraint for optimization here is that the algorithm only requires the surface to be represented 1:1, and therefore the internal geometric space can be ignored (and it is theoretically optimal to achieve such a result, as printing internal geometry is potentially useless on the final assembly). This incentivizes the selection of seed blocks only in the boundary set. Achieving this is trivial: if the block is an internal block, find the nearest boundary block and use that instead.

\begin{figure}
    \includegraphics[width=\columnwidth]{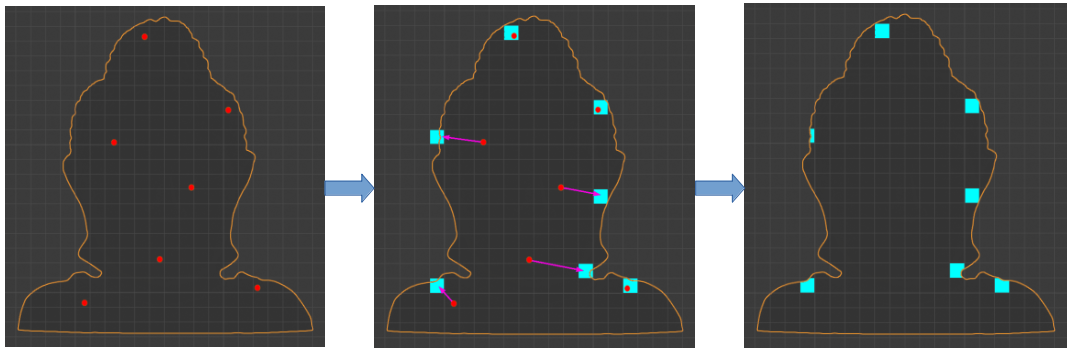}
    \captionof{figure}{2D cross-sectional demonstration of how the initial "seed" blocks are determined. \textit{Left}: Cluster centroids determined from \textit{k}-means++ on vertices; \textit{Middle}: Association of cluster centroids to nearest surface boundary boxes; \textit{Right}: Output initial blocks in cyan.}
    \label{fig:initialblocks}
\end{figure}

The implementation of this is trivial. Using the vertices as data points, apply k-means++ as formally described until converged, and then designate as seed blocks any blocks which enclose a resultant cluster centroid. Find the nearest boundary block if the block is internal. An example of this process can be seen in \textit{Figure \ref{fig:initialblocks}}.

\subsection{Block Growth}
With the initial seed blocks decided, the algorithm can now proceed to the core of the logic this research seeks to implement: growing the blocks to achieve the 1:1 coverage of the model surface in a manner that is optimal for parallel printing.

\begin{figure*}[ht]
    \includegraphics[width=\textwidth]{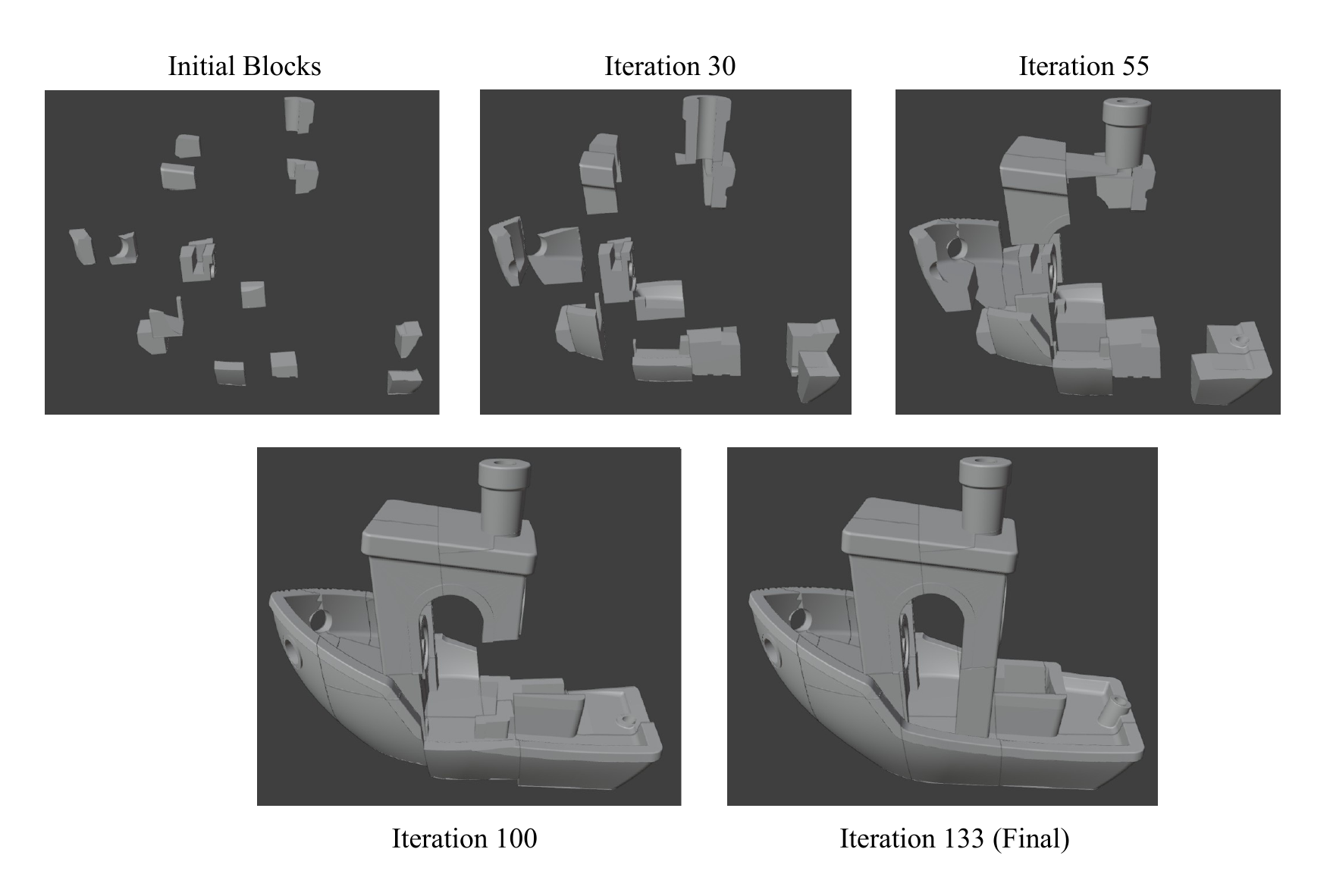}
    \captionof{figure}{Demonstration of the block growth phase on \textit{3DBenchy}.}
    \label{fig:growthiterations}
\end{figure*}

Block growth is performed in an iterative process that takes place in serial. In each iteration, the best block is selected for growth along a specific axis according to an objective function, and growth is performed by expanding the axis-aligned bounding box by a single grid unit in that axis. For example, growing a box, 2x3x4, upward in the Y axis produces a box, 2x4x4, with another "layer" of cells added on the top. In practice, this results in a process that can be seen in \textit{Figure \ref{fig:growthiterations}}. Running this operation in parallel was the original design, but the results of the objective function change as blocks expand, which have implications on subsequent block growth, and led to suboptimal decomposition outcomes.

As a result of using axis-aligned bounding boxes as our blocks, this constrains growth to be only along the base axes, which makes enumerating possible growth options far more definite as there are only 6 directions to consider: +X, -X, +Y, -Y, +Z, -Z.

\begin{figure}[h]
    \includegraphics[width=\columnwidth]{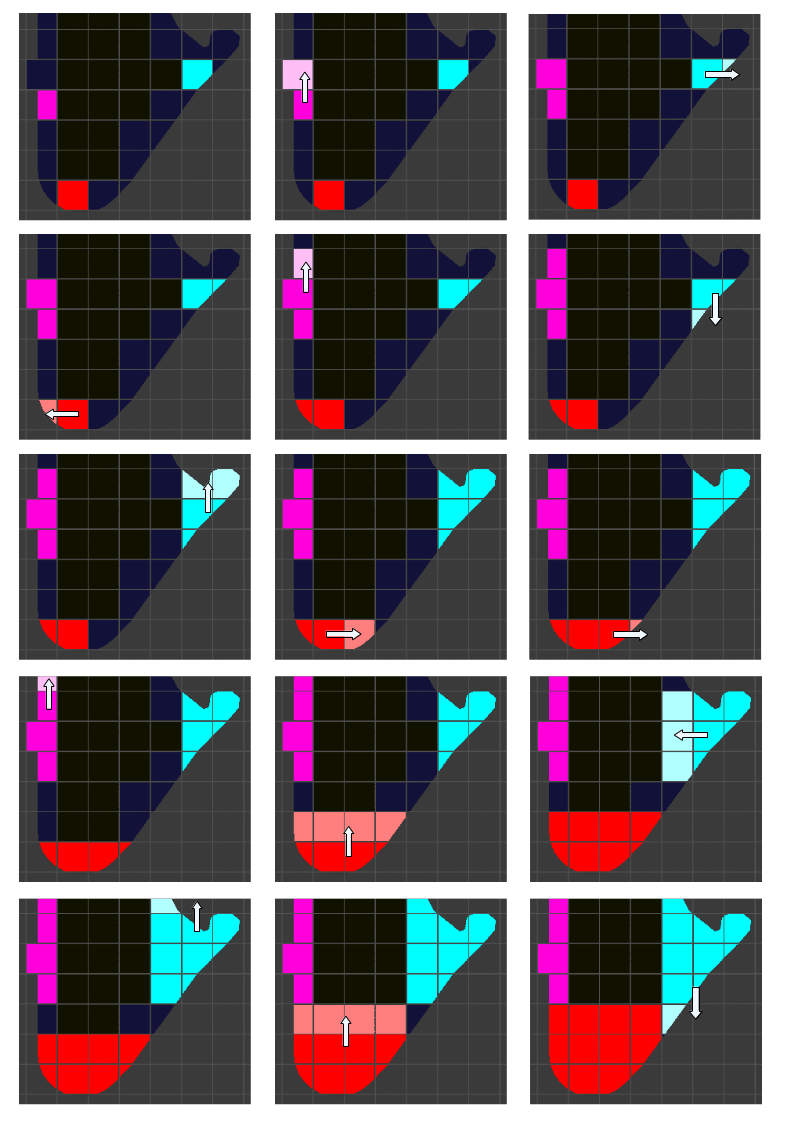}
    \captionof{figure}{2D cross-sectional worked example of the block growth on a section of a model. Pink, red, and cyan seed blocks. Growth iterations depicted with arrows showing direction and highlighted regions showing new regions.}
    \label{fig:blockgrowth}
\end{figure}

\subsubsection{Objective}
The workhorse of the algorithm is the objective function. This drives and directs the block growth process by quantifying as a single number how good or bad the expansion of a block in a specific direction is, allowing for every direction of every block to be enumerated, scored for fitness and ranked.

The scoring follows a simple pattern that is best seen as computing the "cost" of growth. Lower scores are less preferable than higher scores, except for scores $\leq 0$ which indicate that the algorithm should bypass these options.

In formulating this function, there are effectively three different types of variables that are incorporated:

\begin{itemize}
    \item \textbf{Hard Constraint:} These variables incorporate requirements that cannot be violated whatsoever. These effectively prohibit actions that would result in parts that cannot be sanely printed. The violation of any hard constraint forces the objective to -1.
    \item \textbf{Soft Constraint:} These variables model characteristics of growth that should be disincentivized. Unlike the hard constraints, however, they can be tolerated, but will impose penalties. These increase the score of the objective.
    \item \textbf{Incentive:} These variables model characteristics of growth that should be incentivized. These decrease the score of the objective.
\end{itemize}

The objective can be represented as the following expression, where $S(M)$ is a scoring function calculating the score of a given part $M$ (as expanded in the direction $D$):

\begin{align}
    S(M,D) = 
    \begin{cases}
        |P(M) + O(M)| * S_{prox}, \quad & S_{hard}(M) \neq -1\\
        -1, \quad & S_{hard}(M) = -1 \\
    \end{cases}
\end{align}

This expands with the following expressions. $S_{hard}(M)$ here represents the "hard constraints", outputting -1 if any constraints are violated:

\begin{align}
    S_{hard}(M) = 
    \begin{cases}
        -1, \quad & OutOfBounds(M) \\
        -1, \quad & \neg FitsPrinter(M) \\
        -1, \quad & M_{new} = \emptyset \\
        -1, \quad & \exists m \in M_{new} : m_{parents} \neq \emptyset\\
    \end{cases}
\end{align}

$S_{prox}(M)$ determines the "proximity score" for $M$, based on the $L_1$ distance function. $N$ here is the set of all parts. This softly guides block growth away from other regions, incentivizing a more even distribution of the blocks across the model.

\begin{align}
    S_{prox}(M) = \min_{\forall n \in N : n \neq M} dist(M,n)\\
    dist(M,n) = L_1(M_{centroid},n_{centroid}) - (M_{size}+n_{size})
\end{align}

Here, $P(M)$ computes the approximate printing score of $M$. $O(M)$ computes the overhang score, and $Directions$ is the set of all cardinal directions; $orientedOverhangingArea$ is a function that determines the overhang area for $M$ for a given orientation, with the expression choosing the orientation that has the lowest overhanging area. These are both penalties: overhang area and assumed printing speed should be minimized.

\begin{align}
    P(M) = Speed_{infill} * M_{volume} + Speed_{shell} * M_{surfArea}\\
    O(M) = \min_{\forall d \in Directions}(orientedOverhangingArea(M, d))
\end{align}

$P(M)$ is relatively simplistic compared to alternative, more accurate approaches \cite{Medina:2019}, but as previously mentioned, this captures the core 3D printing time cost variables \cite{Raise3D:2024} in a fast heuristic.

\subsubsection{Stopping Condition}
Block growth continues to iterate until certain conditions are met. 

\begin{itemize}
    \item Ideally, the blocks grow to a point of perfectly representing the surface of the model, but this is unlikely to happen; nonetheless, this is tested by halting the growth loop if no "unassigned boundary cells" remain.
    \item Otherwise, the growth loop ends if no further viable growth operations can be determined.
\end{itemize}

The latter point usually triggers when blocks have grown such that any further growth into unassigned regions would result in overlaps with other blocks, which violates the 1:1 representation requirement.

Initially, an alternative strategy was trialed which involved allowing blocks to overlap until all unassigned boundary cells were assigned, and then resolving these overlap "conflicts" using combinations of shrink-expand operations. Major issues were found using this strategy that could not be solved with different approaches. This approach is detailed in \textit{Alternatives} (\textit{Section \ref{alternatives}}).

Consequently, in this implementation, the block expansion tends to stop with unassigned regions still remaining. This requires another step, as follows in the next subsection.

\subsection{Conflict Resolution}
\begin{figure}
    \includegraphics[width=\columnwidth]{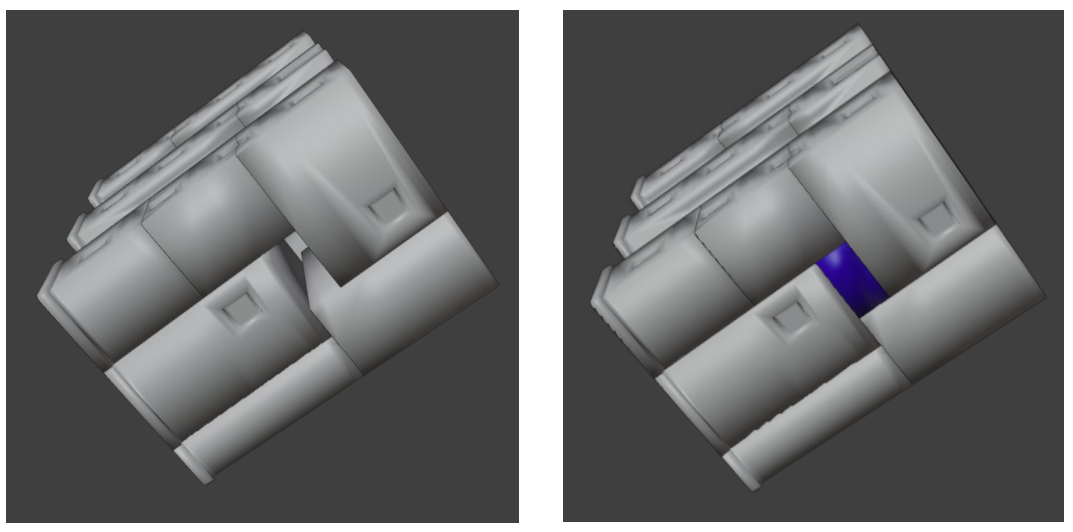}
    \captionof{figure}{Left: Model with a conflict. Notice that none of the surrounding boxes can expand in a way that fills the void without introducing overlaps. Right: New block (blue) created to fill the void.}
    \label{fig:voidfill}
\end{figure}

At this point in the algorithm, the blocks have been expanded to cover as much of the surface as they can without introducing overlaps. Assuming this expansion covered the surface entirely at this point, then conflict resolution is unnecessary and this step is skipped, but as stated previously, it was found that this is rarely the case in practice. This will typically appear as the model being partially covered by blocks, but with "void regions" of the surface boundary between them, as can be seen in the left image of \textit{Figure \ref{fig:voidfill}}.

A baseline strategy was employed which attempted to fill these void regions through the creation of new blocks. This strategy involved the implementation of a method that iterated through unassigned boundary cells, creating seed blocks from them and attempting to grow them until they cannot be grown any further without introducing overlaps, and repeating until the 1:1 representation requirement is attained \textit{(see Algorithm \ref{alg:conflictresolve})}.

\begin{algorithm}
    \caption{Conflict resolution through void filling logic.}
    \label{alg:conflictresolve}
    \begin{algorithmic}		

        \Function{GetDiscreteEmptyRegions}{\textit{grid}, \textit{numFreePrinters}}
            \State $\textit{emptyRegions} \gets \emptyset$
            \State $\textit{iteration} \gets 0$
            \While{\textit{iteration} $<$ \textit{numFreePrinters}}
                \State $\textit{boundaryCell} \gets \textit{NextUnassignedBoundaryCell()}$
                \If{\textit{boundaryCell} $\neg \emptyset$}
                    \State $\textit{cuboid} \gets \textit{CreateUnitCuboidAt(boundaryCell)}$
                    
                    \While{\textit{CanExpandWithoutOverlap}}
                        \ForAll{\textit{direction} $\in$ \textit{XYZDirections}}
                            \If{CanExpandInDirection(\textit{cuboid}, \textit{direction})}
                                \State ExpandInDirection(\textit{cuboid, direction})
                            \EndIf
                        \EndFor
                    \EndWhile
                    \State \textbf{add} \textit{cuboid} \textbf{to} \textit{emptyRegions}
                \EndIf
                \State $\textit{iteration} = \textit{iteration} + 1$
            \EndWhile
            \State \Return \textit{emptyRegions}
        \EndFunction

\\

        \Function{AssignMeshBoxesFrom}{\textit{parent}, \textit{gridCells}, \textit{emptyRegions}}
            \ForAll{\textit{emptyRegion} $\in$ \textit{emptyRegions}}
                \ForAll{\textit{gridCell} $\in$ \textit{gridCells}}
                    \If{\textit{gridCell} \textbf{inside} \textit{emptyRegion}}
                        \State \textbf{add} \textit{gridCell} \textbf{to} \textit{emptyRegion}
                    \EndIf
                \EndFor
                \State $\textit{meshBox} \gets ClipFromMesh(\textit{parent, emptyRegion})$
                \State \textbf{add} \textit{meshBox} \textbf{to} \textit{meshBoxes}
            \EndFor

            \State \Return \textit{meshBoxes}
        \EndFunction
    \end{algorithmic}
\end{algorithm}

The main advantages of this approach are its relative simplicity, speed, and typically adding smaller blocks that do not impede the min-max performance (which will still be the print time of the largest block). The primary drawback of this approach is that it involves the creation of supernumerary blocks, which violates the printer quantity constraint specified by the user. However, in cases where there are less blocks needed than this quantity, this gives some leeway in adding additional blocks.

\subsubsection{Alternatives} \label{alternatives}

Alternative strategies were formulated to try to improve these characteristics. The main strategy of note involved identifying empty regions and, for each region, enumerating all possible expansion operations of adjacent blocks, and then all possible shrink operations of the blocks that are brought into conflict through overlaps. The best one in terms of minimizing maximum block size while maintaining the perfect fit would be taken.

In practice, this approach revealed itself as both overengineered and flawed:

\begin{itemize}
    \item It worked appropriately on only a shockingly small number of use cases, as chains of conflicts often echoed into complex configurations.
    \item Expanding neighboring blocks often served to actually reduce the min-max performance, increasing the propensity of the algorithm towards increasing the sizes of the largest blocks even further.
\end{itemize}

These drawbacks, combined with time constraints that impair the practicality of fixing the issues involved with the implementation, caused the approach to be abandoned in favor of the simple void fill approach. The deciding factor for this came with the following step of the implementation, which worked better with this strategy.

\subsection{Metaheuristic}
By this point, the algorithm pipeline functionally works to produce valid decompositions of input models according to the objective metric. This final section of the method details high-level improvements to the algorithm that substantially improves its performance, albeit at the expense of significantly more computation time.

\begin{figure}[h]
    \includegraphics[width=\columnwidth]{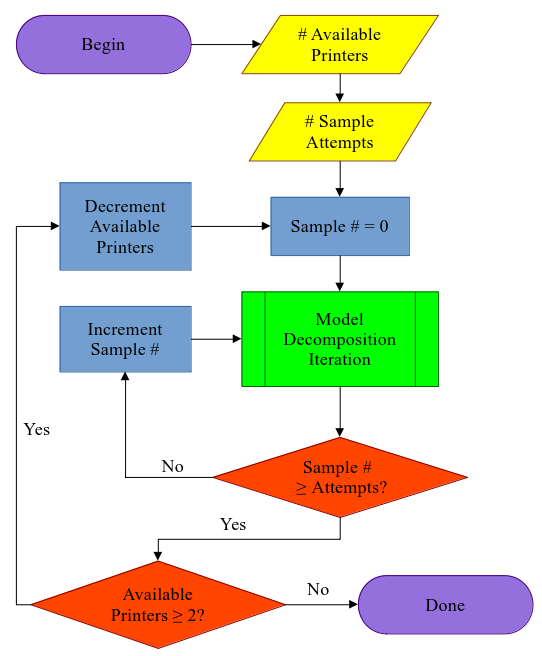}
    \captionof{figure}{Flow chart depicting the loops of the metaheuristic, and how they intersect with the inner model decomposition process iteration (i.e. all of the prior steps in this paper section).}
    \label{fig:metaheuristic}
\end{figure}

A variety of observations can be made about the process thus far:

\begin{itemize}
    \item Sometimes, sufficient parallelization can be achieved with fewer printers.
    \item Spare printers enables the \textit{Conflict Resolution} stage to be more likely to result in a printable output.
    \item While the core of the algorithm is fully deterministic, the k-means++ nature of the initialization step introduces a stochastic element.
\end{itemize}

Taking all of these together, it can be assumed that each run can be somewhat different, unless the clustering is strongly defined. With the extra processing power available, this can be utilized to the algorithm's advantage: multiple different attempts can be made, and the best one chosen.

This can then be expanded even further: repeating the process and reducing the quantity of printers supplied to the \textit{Initialization} and \textit{Block Growth} stages allows for further attempts with increased availability of printers to the \textit{Conflict Resolution} stage. By then eliminating invalid results and choosing the best result, this expanded step strikes a balance between ensuring a printable configuration, searching for an optimal configuration in print time, and seeking to reduce the number of printers needed (if possible).

This can be represented as the application of the following expression:

\begin{align}
    i_{best} = \arg \min_{S(i) \forall i \in I},\\
    where S(i) = \max_{P(r) \forall r \in R(i)}
\end{align}

where $I$ is the set of all outputs of all iterations, $P(r)$ is the printing score of a given part $r$, $R(i)$ is the set of parts for a given iteration $i$, $S(i)$ is the parallel score of iteration $i$, and $i_{best}$ represents the index of the best iteration selected as the final output of the entire algorithm.

Implementing this metaheuristic takes the form of two loops: one outer loop that reduces the initially supplied printers each iteration, and one inner loop that attempts different clustering configurations per iteration (\textit{see Figure \ref{fig:metaheuristic}}).

\section{Results}
The algorithm was evaluated by comparison with the two principle alternative approaches in the literature: \textit{An Algorithm For Partitioning Objects Into A Cube Center And
Segmented Shell Covers For Parallelized Additive Manufacturing}, and \textit{Symmetry-Based Decomposition For Optimised Parallelisation In 3D Printing Processes} \cite{Hatton:2023}. 

The native implementation of the \textit{Cube Skeleton Segmented Shell} algorithm was unable to be acquired. A faithful rendition of this was implemented in C++ using \textit{CGAL} to the formal specification as published. Some minor errors in the author's paper were discovered fixed in the implementation. Some minor optimizations were also made, which will have improved computation time, but will have had no effect on the final decomposition.

\subsection{Experimental Design}
In order to achieve a sufficient yet practical evaluation, a sample set of geometry is selected and each algorithm is run across it with certain fixed parameters that represent reasonable optimal cases. The results are subjected to a primarily quantitative method, with qualitative feedback weakly informing the final conclusions.

Contrasting with the \textit{Symmetry-Based Decomposition} experimental design, only larger models are appropriate, as the results indicated little performance gain from the decomposition of smaller models that can already be printed on a single printer, and this algorithm was developed with that feedback. This also fits the sample pattern of the other control algorithm, which focused on larger objects.

A representative set of models was chosen to test the algorithms across various domains. These included variations in complexity (simple to complex), topology, and symmetry. This provides a wide range for assessing the performance of the algorithms across various properties, while also combining to make a good average assessment. It is expected that \textit{Parallelobox} will perform better primarily on more complex objects, where the benefits of AABBs should better tolerate complex features over the other algorithms, with secondary benefits in symmetric cases due to the early optimization stage.
These models were selected from a variety of sources. Some are standard test objects common to the field. Some were downloaded from \textit{Thingiverse}. Some were derived from the \textit{Thingi10K} data set, using \textit{Symmetry-Based Decomposition}'s sample set of appropriate and representative models. "\textit{Brain Left}" was a \textit{NIfTI} MRI image of a friend's left hemisphere, converted into an STL file and given with permission to use. The sample set and their qualities are specified in \textit{Table \ref{tbl:models}}.

\begin{table*}[ht]
\centering
\begin{tabular}{|l|c|c|c|}
\hline
\multicolumn{1}{|c|}{\textbf{Model}} & \textbf{Symmetry} & \textbf{Complexity} & \textbf{Topology} \\
\hline
\textbf{3DBenchy}                               & Symmetric & Complex & Non-Convex \\
\hline
\textbf{Utah Teapot}                            & Symmetric & Complex & Non-Convex \\
\hline
\textbf{Stanford Bunny}                         & Asymmetric & Simple & Convex \\
\hline
\textbf{Snail}                                  & Asymmetric & Simple & Convex \\
\hline
\textbf{Large Building}                         & Symmetric & Simple & Convex \\
\hline
\textbf{Brain Left}                             & Asymmetric & Complex & Non-Convex \\
\hline 
\textbf{797990 \textit{("Trousers")}}           & Symmetric & Simple & Convex \\
\hline
\textbf{78224 \textit{("Hole Pipe")}}           & Symmetric & Complex & Non-Convex \\
\hline
\textbf{500094 \textit{("Sparse Lattice")}}     & Symmetric & Complex & Convex \\
\hline
\textbf{1053374 \textit{("Strange Washer")}}    & Asymmetric & Simple & Non-Convex \\
\hline
\textbf{77951 \textit{("Side Plate")}}          & Asymmetric & Simple & Convex \\
\hline
\textbf{41084 \textit{("Mini Refinery")}}       & Asymmetric & Complex & Convex \\
\hline
\textbf{51329 \textit{("Unusual Vase")}}        & Asymmetric & Simple & Non-Convex \\
\hline
\textbf{96457 \textit{("F-177 Nighthawk")}}     & Symmetric & Complex & Convex \\
\hline
\end{tabular}
\caption{Qualitative properties of the sample models. \textit{Thingi10K} sample models have their human-appropriate assigned names indicated next to their number.}
\label{tbl:models}
\end{table*}

For this algorithm, the parameters were fixed across all sample models. The target printer's specifications were stored as a .ini file for the algorithm to use, and each model was run using the same settings and arguments (seen in \textit{Table \ref{tbl:settings}}), with the number of printers varying per experiment to test the performance of the algorithm. 

There are also theoretical internal algorithmic parameters, such as scale factors for overhang and the $l_1$ distance to adjust their penalties in the objective. These were removed (e.g. modeled as a scale factor of 1.0) in the acquisition of these results to capture a baseline performance.

\begin{figure}
    \includegraphics[width=\columnwidth]{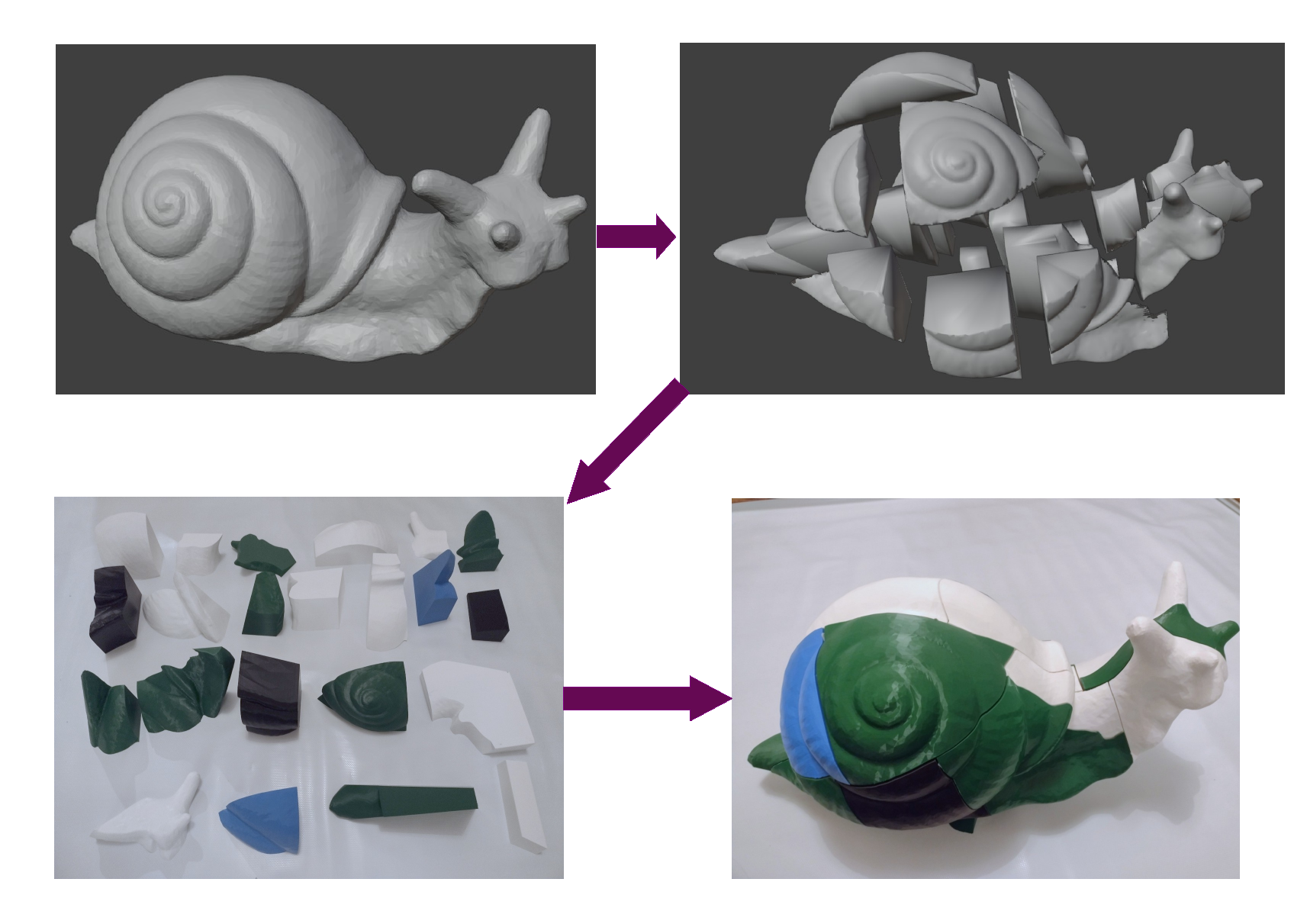}
    \captionof{figure}{Demonstration of the process from model to assembly. Top Left: Input model. Top Right: Exploded view of the algorithm's output partition set. Bottom Left: The printed parts of the partition set. Bottom Right: Assembled model.}
    \label{fig:snassembly}
\end{figure}

\begin{table}[H]
\centering
\begin{tabular}{|l|c|}
\hline
\multicolumn{1}{|c|}{\textbf{Setting}} & \textbf{Value} \\
\hline
\textbf{Infill}             & 0.05 \textit{(5\%)} \\
\hline
\textbf{Sample Tries}       & 3 \\
\hline
\textbf{Skip Symmetry}      & No \\
\hline
\textbf{Granularity}        & Very Fine \textit{($\sim15^3$ cells)} \\
\hline
\textbf{Overhang Tolerance} & 1\textdegree \\
\hline
\textbf{Printer Volume}     & 250 x 250 x 250 mm \\
\hline
\textbf{Shell Speed}        & 20 mm/s \\
\hline
\textbf{Infill Speed}       & 20 mm/s \\
\hline
\textbf{Number of Printers} & \makecell{10, 20, 30, 40, 50, \\60, 70, 80, 90, 100} \\
\hline
\end{tabular}
\caption{\textit{Parallelobox} settings and arguments used for all sample models.}
\label{tbl:settings}
\end{table}

\subsection{Data Acquisition}
The resulting prints were largely simulated due to resource constraints. Prior experience with the printers, with corroborated findings in prior research \cite{Hatton:2023}, suggests that the slicer's computed timings translate reasonably well into real time on the scales dealt with here. Some results were physically printed and assembled to validate the algorithm in real world space, but the large models combined with the resource constraints impeded the quantity of these to only select models.

All models were sliced into GCode using \textit{Ultimaker Cura}, for printing on \textit{Ultimaker 2+ Extended} 3D printers using PLA. Slicing parameters were kept the same across all models and selected to reflect reasonable consumer use, to ensure valid results. Most settings were kept as Cura's defaults for fast settings, with these settings being notable selections or alterations:

\begin{table}[H]
\centering
\begin{tabular}{|l|l|} 
\hline
\textbf{Infill Density}     & 5\%            \\ 
\hline
\textbf{Infill Pattern}     & Zig-Zag        \\ 
\hline
\textbf{Layer Height}       & 0.25 mm        \\ 
\hline
\textbf{Support Structures} & \makecell{Overhang Only \\ \textit{(45° tolerance)}}  \\ 
\hline
\textbf{Plate Adhesion}     & Brim \textit{(8.0 mm)}  \\
\hline
\end{tabular}
\caption{\textit{Ultimaker Cura} slicing properties utilized across all sample models.}
\label{tbl:slicingproperties}
\end{table}

Specifically, an automated process was employed in which the algorithms were run on the sample models in batch, after they were first run through \textit{MeshFix}, an implementation that automatically fixes common meshing problems such as self-intersections \cite{Attene:2010}. The output parts were then subjected to batched post-processing, which involved automatically centering and rotating the parts for optimal 3D printing using \textit{Tweaker-3} \cite{Schranz:2021}. Cura's \textit{CuraEngine} back-end was interfaced with by the command line to automate the slicing, and the estimated printing times were extracted from the output log. The parallel printing time for each model was computed from these data using a shell script. 

\begin{figure}[H]
    \includegraphics[width=\columnwidth]{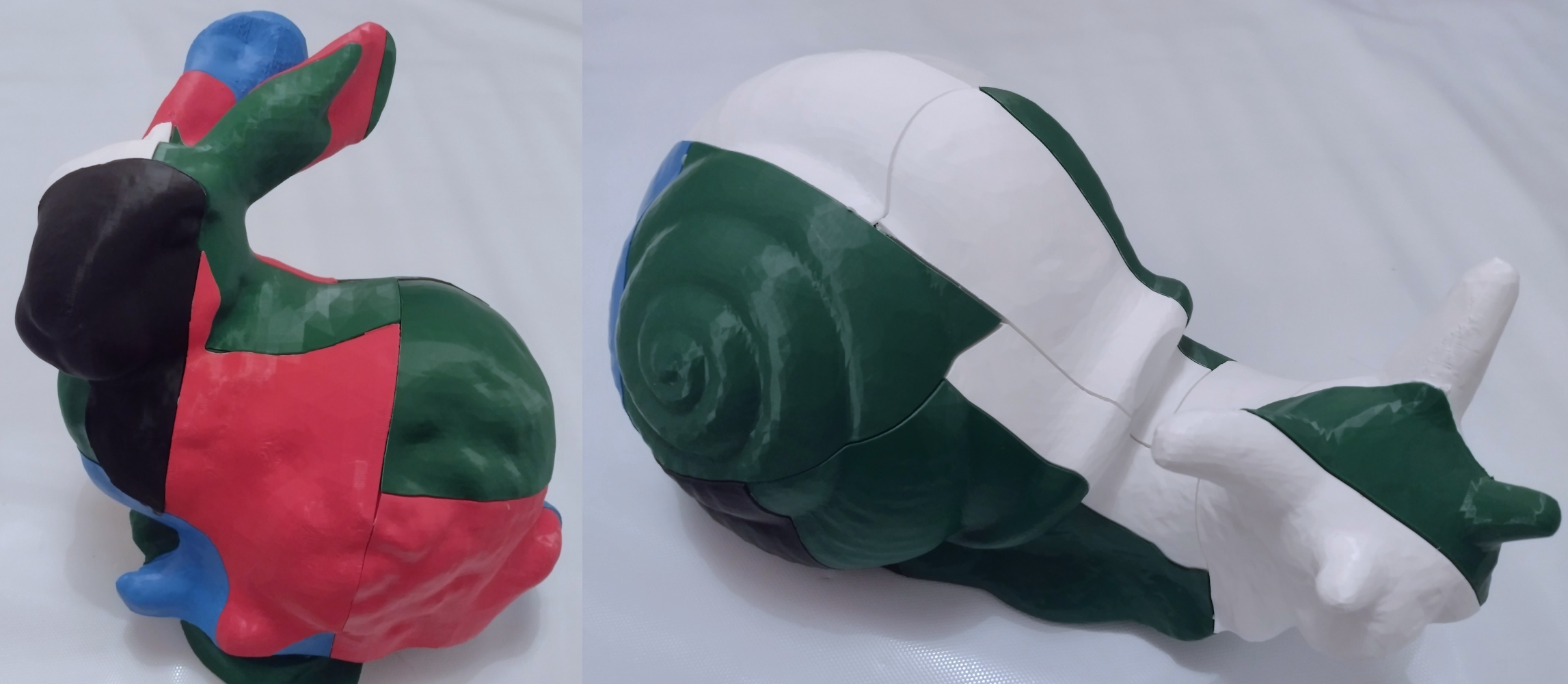}
    \captionof{figure}{The two models produced in the real world. The Oversized \textit{Stanford Bunny} on the left took 45 minutes to assemble. The Oversized \textit{Snail} on the right took 42 minutes to assemble. Both were assembled from an unsorted collection; timings also included adhesive applications.}
    \label{fig:examplemodels}
\end{figure}

\subsubsection{Equipment Setup}

These models were then printed on \textit{Ultimaker 2+} and \textit{Ultimaker 2+ Extended} printers due to availability and to minimize any confounding variables. 

The printers featured \cite{MakerBot:2022}:

\begin{itemize}
    \item 2.85mm PLA filament, fed by bowden tube.
    \item A 0.4mm nozzle, heated to 200$^{\circ}$C.
    \item Glass build plate, heated to 50$^{\circ}$C.
    \item Build volume of size 223 mm x 223 mm x 205 mm.
    \item Gantry-based 2D extruder motion, with 12.5$\mu$m positioning precision, with the bed on a vertical motion Z axis at 5.0$\mu$m positioning precision.
    \item Print speed of 30-300 mm/s, and travel speed of 30-350 mm/s.
\end{itemize}

\subsection{Analysis}

\begin{figure*}[ht]
    \includegraphics[width=\textwidth]{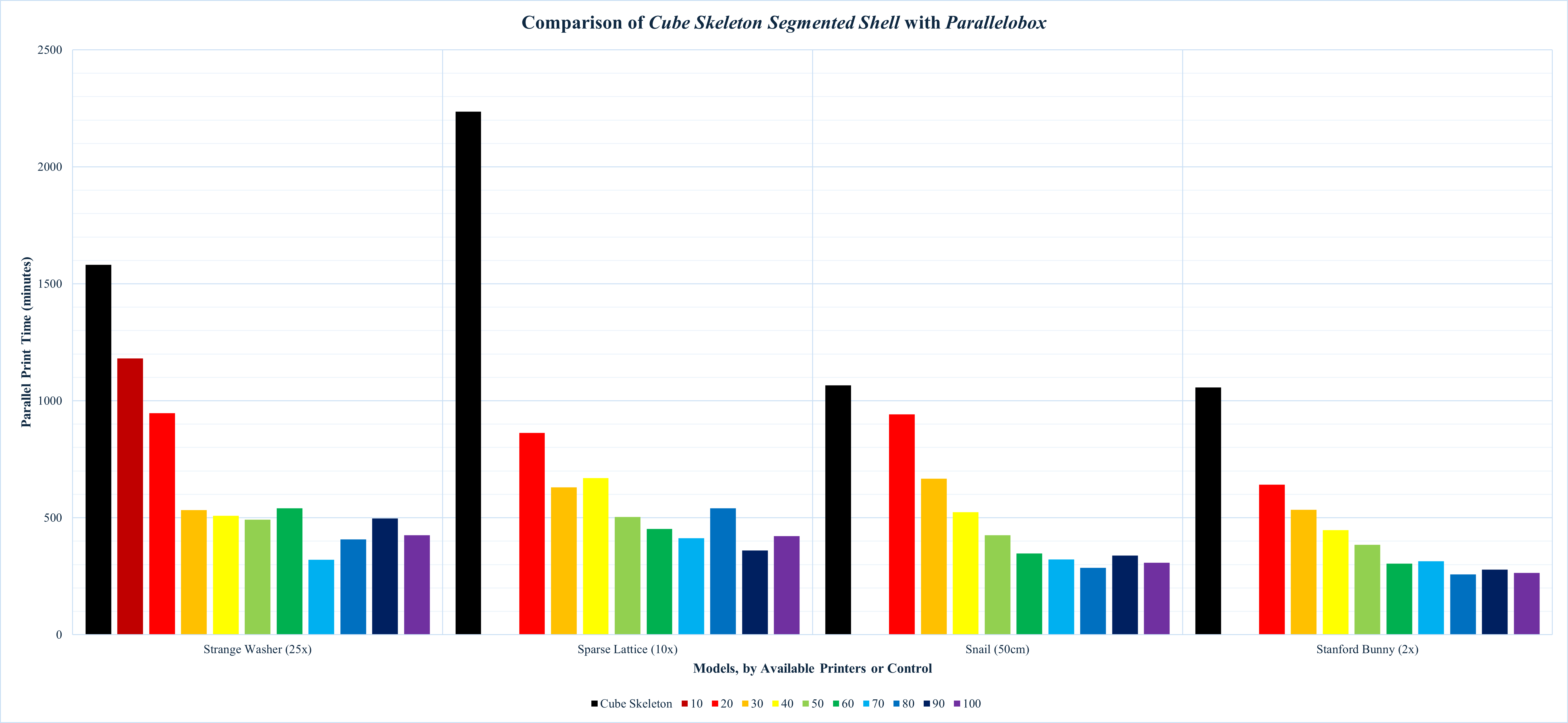}
    \captionof{figure}{Sample comparison of parallel printing times of \textit{Cube Skeleton Segmented Shell} against \textit{Parallelobox} across each sample model. The black bars are \textit{Cube Skeleton}'s result, the rainbow bars are \textit{Parallelobox}'s result with different quantities of available printers. See \textit{Appendix A} for the complete data.}
    \label{fig:segshellvsparallelobox}
\end{figure*}

\begin{figure*}[ht]
    \includegraphics[width=\textwidth]{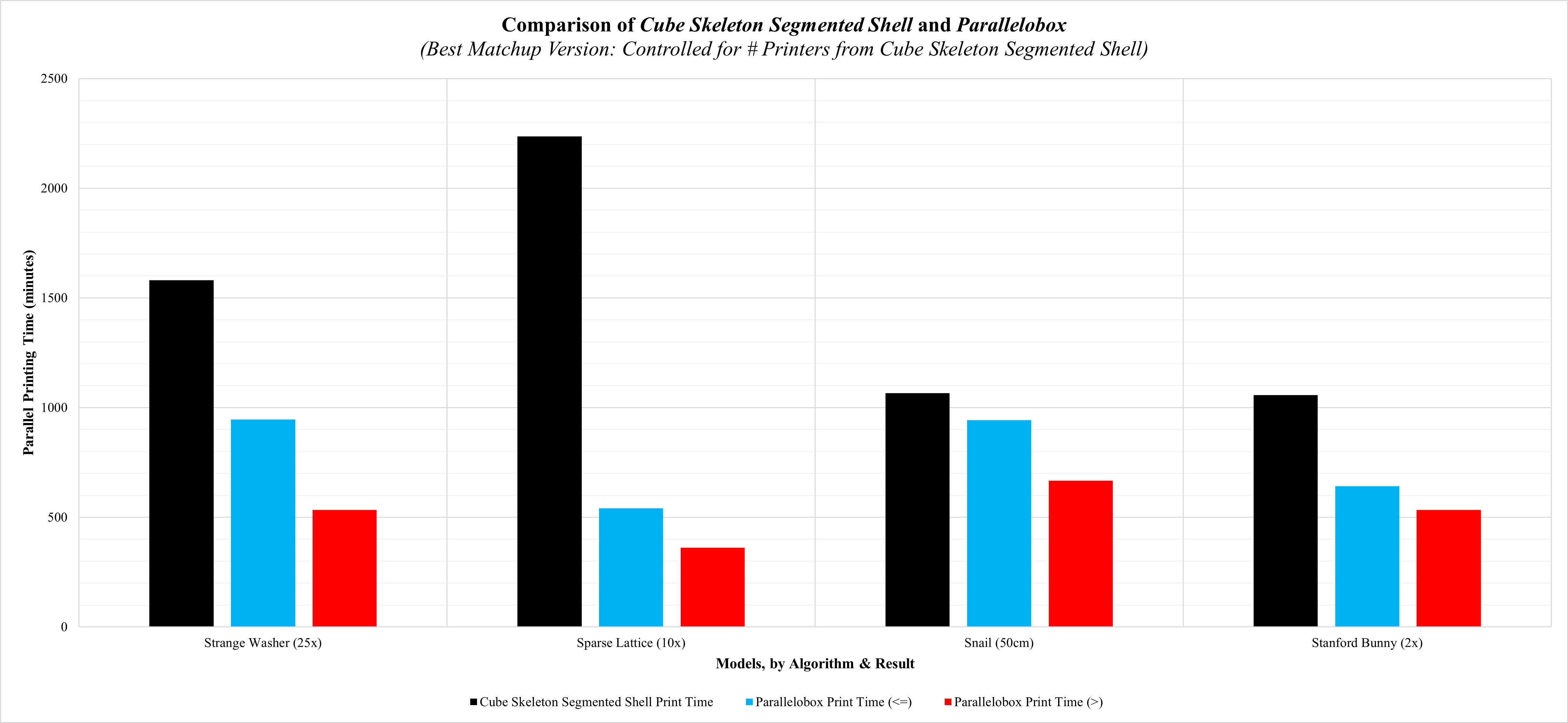}
    \captionof{figure}{Sample "Matchup" comparison of parallel printing times of \textit{Cube Skeleton Segmented Shell} against \textit{Parallelobox} across each sample model. As the former algorithm outputs variable quantities of partitions, this finds the nearest \textit{Parallelobox} results by quantity of partitions. See \textit{Appendix B} for the complete data.} \label{fig:segshellvsparalleloboxmatchup}
\end{figure*}

Comparing \textit{Parallelobox} to \textit{Symmetry-Based Decomposition} is trivial, as both algorithms require a quantity of printers to be specified as "available", which affects the performance of the algorithm, so a direct side-by-side comparison is possible. This is not the case with \textit{Cube Skeleton Segmented Shell}, which has no such input, and outputs any number of partitions for a given model. The single result per model can be compared against \textit{Parallelobox} with a large range of input printers for a good idea of performance. However, in order to ensure as fair a test as possible, a "matchup" scenario can be envisioned. This matchup compares the \textit{Cube Skeleton Segmented Shell} result with the two \textit{Parallelobox} results that most closely bracket it in terms of the number of printers (or a single result, if the number of printers is the same).

In all cases, as a lower printing time is desired, the smaller the result, the better.

\begin{figure*}[ht]
    \includegraphics[width=\textwidth]{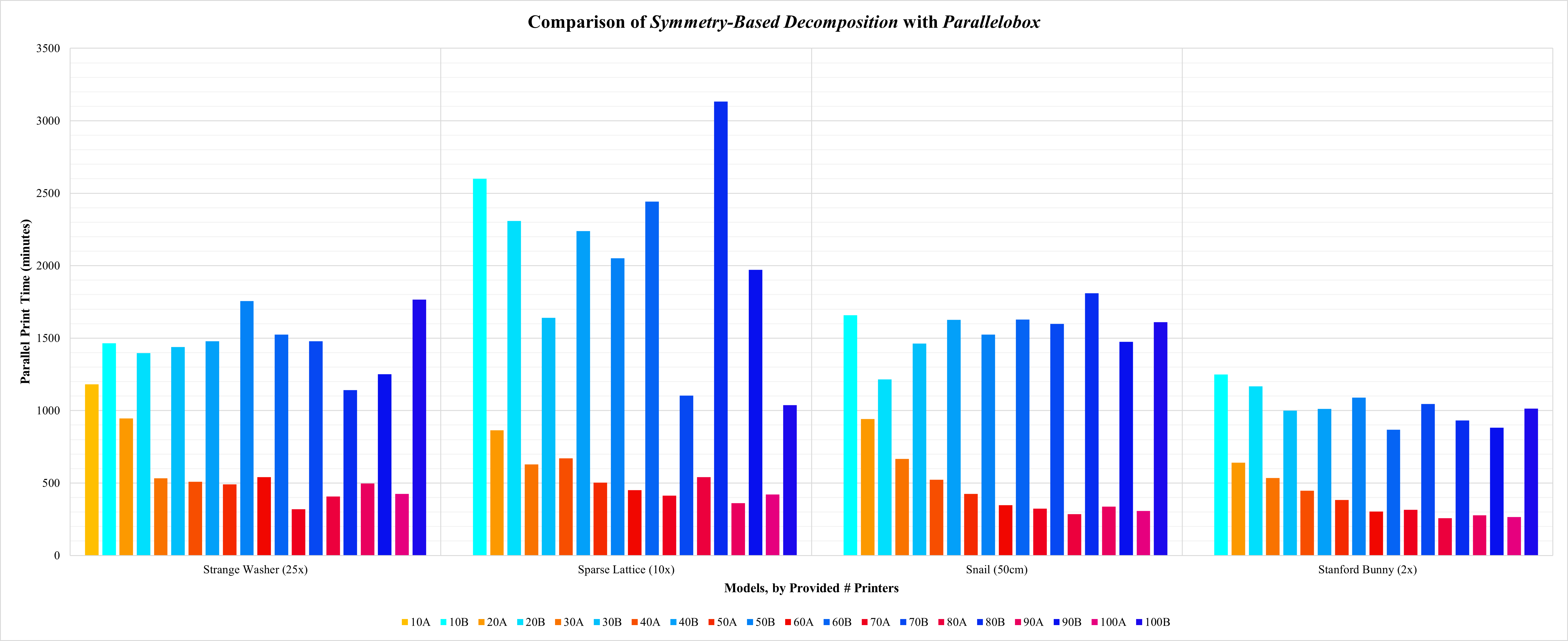}
    \captionof{figure}{Comparison of parallel printing times of \textit{Symmetry-Based Decomposition} (blue hues) against \textit{Parallelobox} (red hues) across each sample model. See \textit{Appendix C} for the complete data.}
    \label{fig:symslicevsparallelobox}
\end{figure*}

Against \textit{Symmetry Slicer}, the results are very apparent: \textit{Parallelobox} all but strictly dominates it. \textit{Symmetry Slicer} provides an idealistic strategy that is unaware of features and support requirements, and while its trivially recursive nature allows it to nearly always produce a valid decomposition, in every match-up, \textit{Parallelobox} produced a superior decomposition set. This superiority became even more pronounced with more complex models and as the parallelization was scaled up with more available printers.

Against \textit{Cube Skeleton Segmented Shell}, the results also speak for themselves. \textit{Parallelobox}'s decomposition set was almost always at least equivalent with \textit{Cube Skeleton Segmented Shell}, but was usually superior. Simpler, smaller, and more convex models pushed the algorithms toward parity, but more complex, larger, and less convex models demonstrated \textit{Parallelobox}'s advantages more strongly, producing a larger disparity in timings, which, as can be seen with \textit{3DBenchy} and \textit{Brain Left}, can be dramatic. Another notable improvement is the ability to specify a set quantity of printers; the fixed size of the output set of \textit{Cube Skeleton Segmented Shell} makes it suboptimal in settings where a fixed number of printers are available, while \textit{Parallelobox} alters its behaviour to match the user's specification. 

These strengths of \textit{Parallelobox} do not come without drawbacks, however. As can be seen in comparison to both \textit{Cube Skeleton Segmented Shell} and \textit{Symmetry Slicer}, the computation times of \textit{Parallelobox} are magnitudes more than either. This reflects what can be predicted from the nature of these algorithms: \textit{Parallelobox} is a complex multi-stage iterative algorithm, whereas the others are more trivial. This complexity provides the algorithm with more contextual awareness of the model's geometry, but this evidently must come at a cost. Fortunately, although the computation time is magnitudes higher, this still does not translate into much compared to the potential printing time savings: \textit{Cube Skeleton Segmented Shell} may be able to run in milliseconds whereas \textit{Parallelobox} requires several minutes to complete, but several minutes is still not a massive expenditure in the context of many hours of parallel printing time savings.

Analyzing \textit{Parallelobox} by itself, it can be noted that the algorithm exhibited the usual properties of idealized parallelization: a rapid improvement in printing time that has diminishing returns as the degree of parallelization is increased. However, it can be observed that it was occasionally unable to produce a decomposition set within the given number of printers. The stochastic nature of many stages of the algorithm also shows up in the results, with a degree of "noise" being evident (resulting in, for example, a higher quantity of printers sometimes producing an inferior decomposition set). Nonetheless, decent performance of the algorithm can be observed on the larger scale of the results, which are most important in assessing the viability of stochastic processes, where inferior configurations \textit{can} happen, but \textit{on average} the results are improved.

\section{Conclusion}

This paper presents an algorithm to decompose models for optimized parallel printing using the placement and growth of axis-aligned bounding boxes as means to partitioning. Axis-aligned bounding boxes enable a relatively low-cost means of producing a coarse decomposition of the input model, which can be augmented by exploiting an initial symmetric cut and rotating to align optimally within the axes. By utilizing an objective function, the growth of these boxes from initial cluster-based seed blocks can be controlled for optimal volume and surface area distribution, while also minimizing support requirements and unnecessary printing of internal spaces.

This paper demonstrates the significant advantages of this approach against two contemporary algorithms in the parallel printing research space. \textit{Parallelobox} performed at least equivalent to, or usually outperformed, these algorithms in terms of the parallel printing time of their decomposition sets, at the cost of some additional computing power requirements. The top-down nature of both \textit{Symmetry-Based Decomposition} and \textit{Cube Skeleton Segmented Shell} made them both ill-equipped for complex geometries, while \textit{Parallelobox}'s bottom-up nature was able to grow the boxes around the features of these complex geometries.

\subsection{Limitations}

Despite the generally positive results found in this paper, there were limitations to both the design of \textit{Parallelobox} and to the experiment.

As demonstrated in the results, \textit{Parallelobox} recurrently failed to produce valid decomposition sets for certain models until a certain number of printers was specified, even though the volume in which to print them was available. This occurred because the growth of the blocks was being too heavily constrained by the objective. The objective function as designed is relatively strict in penalizing qualities such as overhang and proximity, but these are weighted and the weights can be adjusted. These were held constant through the experimentation to keep the results consistent, but this objective could be tweaked for more flexibility.

This limitation also extends beyond the objective's parameters above. The algorithm has many other points of control which were also held constant. As implemented, the program can take many other arguments to control these, such as specifying the granularity of the grid for producing the initial base decomposition. The experiment utilized a fine setting for balance, but a coarser setting would have produced much faster computation times at the cost of reduced parallel printing time performance; conversely, an even finer setting would have improved the printing time even further but at much higher computation time cost.

Assembly time testing was not tested for. Due to constraints in printing, such as inconsistent and poor availability of the printers, only a very select sample of prints could be produced, which were mostly for demonstration. The time to manually assemble the pieces would be interesting to know; as \textit{Symmetry-Based Decomposition} demonstrated, these times increased with higher levels of decomposition, to the point of overriding the printing performance gains. This could have substantial implications for the research, but was unable to be tested for here.

\subsection{Further Work}

As with the other contemporary algorithms compared through the paper, the interfaces between parts are smooth and flat. Reassembling models thus requires the use of adhesives. Other partitioning algorithms exist that have demonstrated the use of interlocking connectors to improve the assembly of printed models, reducing the need for adhesives and producing strong assembled structures \cite{Song:2015} \cite{Song:2017}. This paper did not incorporate these strategies due to the assembly time not being measured for the other algorithms and being of secondary importance here, as well as the added complexities of accounting for interlocking structures and arrangements in an already complex algorithm. Advantages would likely be seen in the assembly stage if interlocking connectors were incorporated into \textit{Parallelobox}.

\textit{Parallelobox} relies on a stochastic method to find an ideal (or often just sufficient) seed box configuration; this was initially selected because certain configurations will result in situations where the axis-aligned boxes cannot grow without overlapping other boxes, and resolving these stalemates analytically proved a complex task. Theoretically, an analytical resolution algorithm could be developed to this end, which would relegate the need for the computationally expensive stochastic method, and also result in the complete algorithm becoming more deterministic.

As with \textit{Symmetry-Based Decomposition}, \textit{Parallelobox} takes the quantity of available printers as input from the user to guide the algorithm. However, it is conceivable that the user either does not know, or does not care, about the quantity of printers. A crude solution here could be the use of a heuristic or other sub-algorithm which could produce a reasonable number of printers for a given model and use that as input for \textit{Parallelobox}. The algorithm itself could also be modified to find an "optimal" amount of printers, which is to some extent performed in this paper's \textit{Results} section by repeating the process with different quantities of printers, but could also be performed by incorporating other logic into it that is currently unknown.

Equally, in all assessed algorithms here, the printers are assumed to be uniform in the set. In other words, \textit{all} available printers have the same characteristics: the same speeds and the same printing volume geometry. This can be common in lab environments, but a variation in printer models is more likely. Adjusting \textit{Parallelobox} to account for variations in printer models is possible, but would require significant effort, both in modifying the data structures of the parts to explicitly reference individual printer instances, but also in balancing the distribution of models across them while maintaining the optimal parallel printing time.

\section{Declarations}

    \begin{itemize}
        \item \textbf{Authors' Contributions}
    
        Concept: H.H.; Design: H.H.; Implementation: H.H.; Experimentation: H.H.; Supervision: U.M. and M.K.; Data analysis and interpretation: H.H.; Writing: H.H.; Reviewing and editing: J.M. and M.K.; Final approval of the article: M.K. and U.M; Publishing: M.K.
    
        \item \textbf{Funding}
    
        This research was part of a PhD project at the University of Hull and consequently was indirectly funded by the institution.
    
        \item \textbf{Conflicts of Interest}
    
        The authors declare no competing interests.
     \end{itemize}


%





\ifCLASSOPTIONcaptionsoff
  \newpage
\fi





\bibliographystyle{IEEEtran}
\bibliography{Bibliography}

\clearpage
\onecolumn

\section{Appendices}

\subsection{Appendix A: Parallelobox vs Cube-Skeleton Segmented Shell Complete Data} \label{appendixA}

\begin{figure*}[hb!]
    \includegraphics[width=0.9\textwidth]{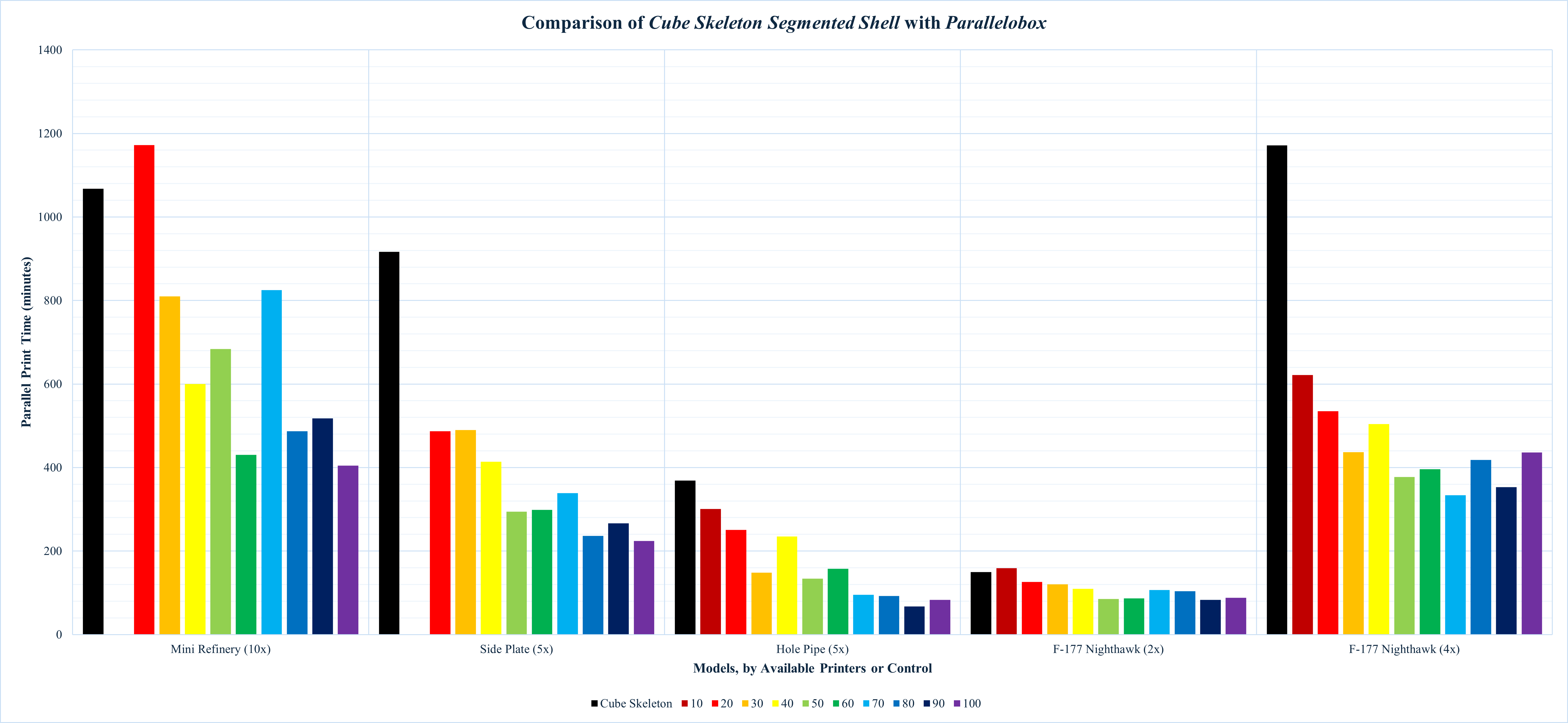}
    \includegraphics[width=0.9\textwidth]{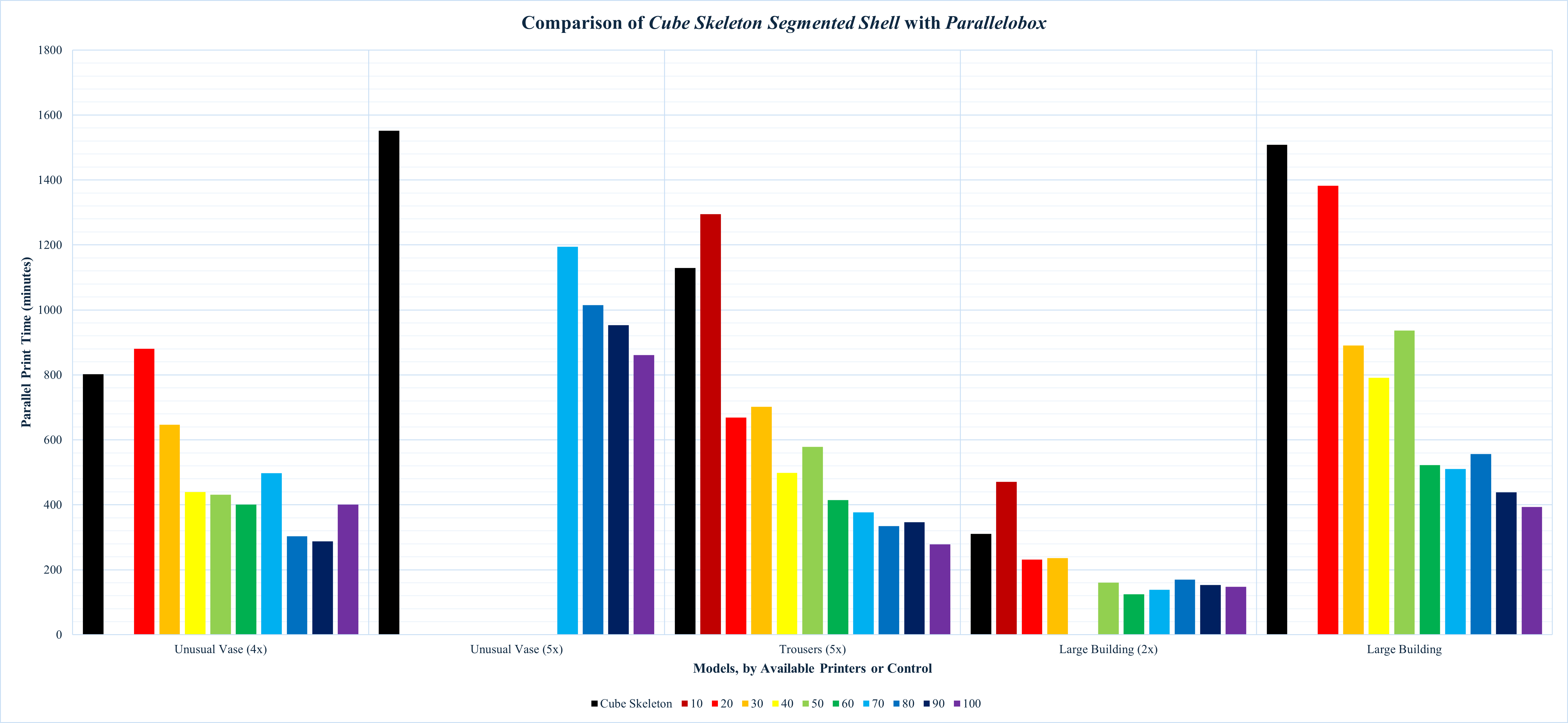}
    \includegraphics[width=0.9\textwidth]{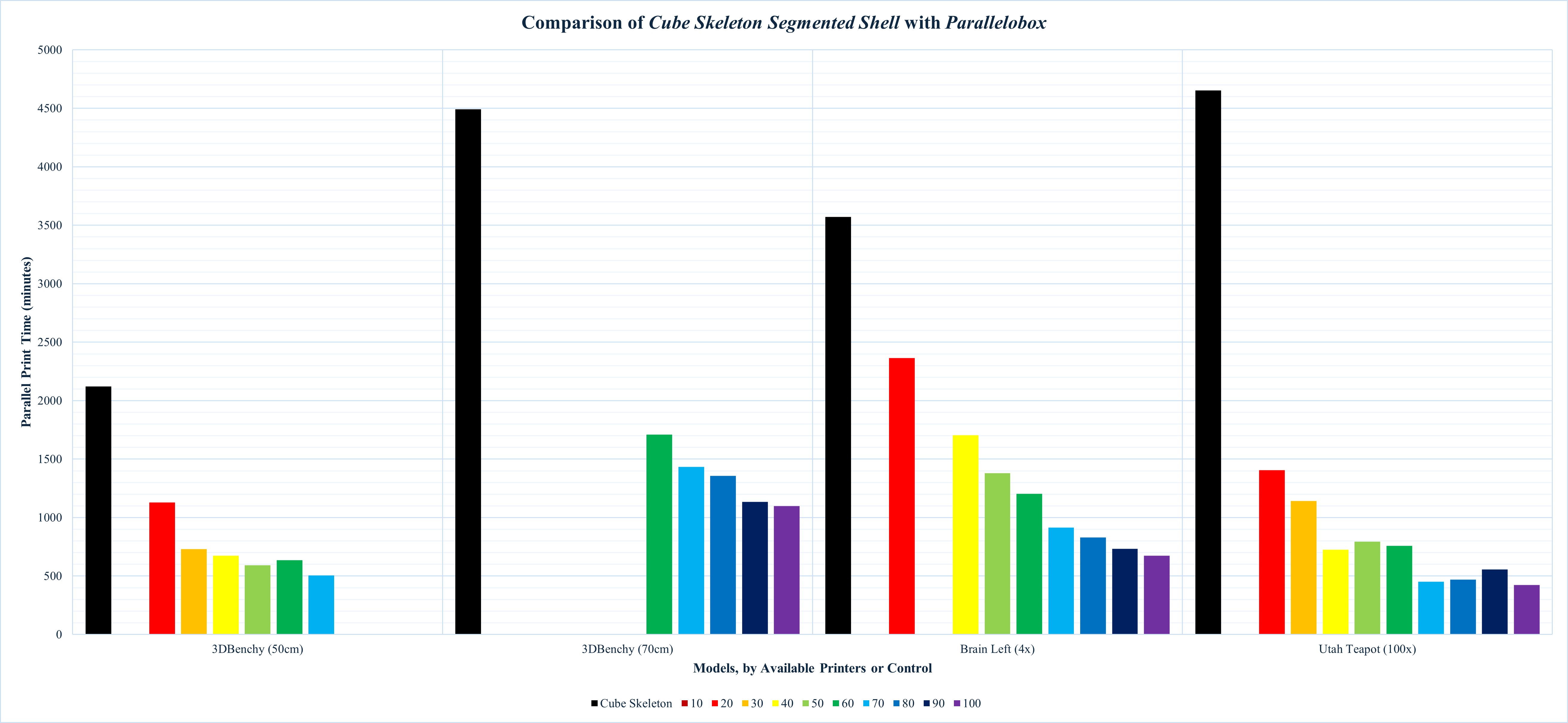}
\end{figure*}

\clearpage

\subsection{Appendix B: Parallelobox vs Cube-Skeleton Segmented Shell Complete Matchup Data} \label{appendixB}

\begin{figure*}[hb!]
    \includegraphics[width=\textwidth]{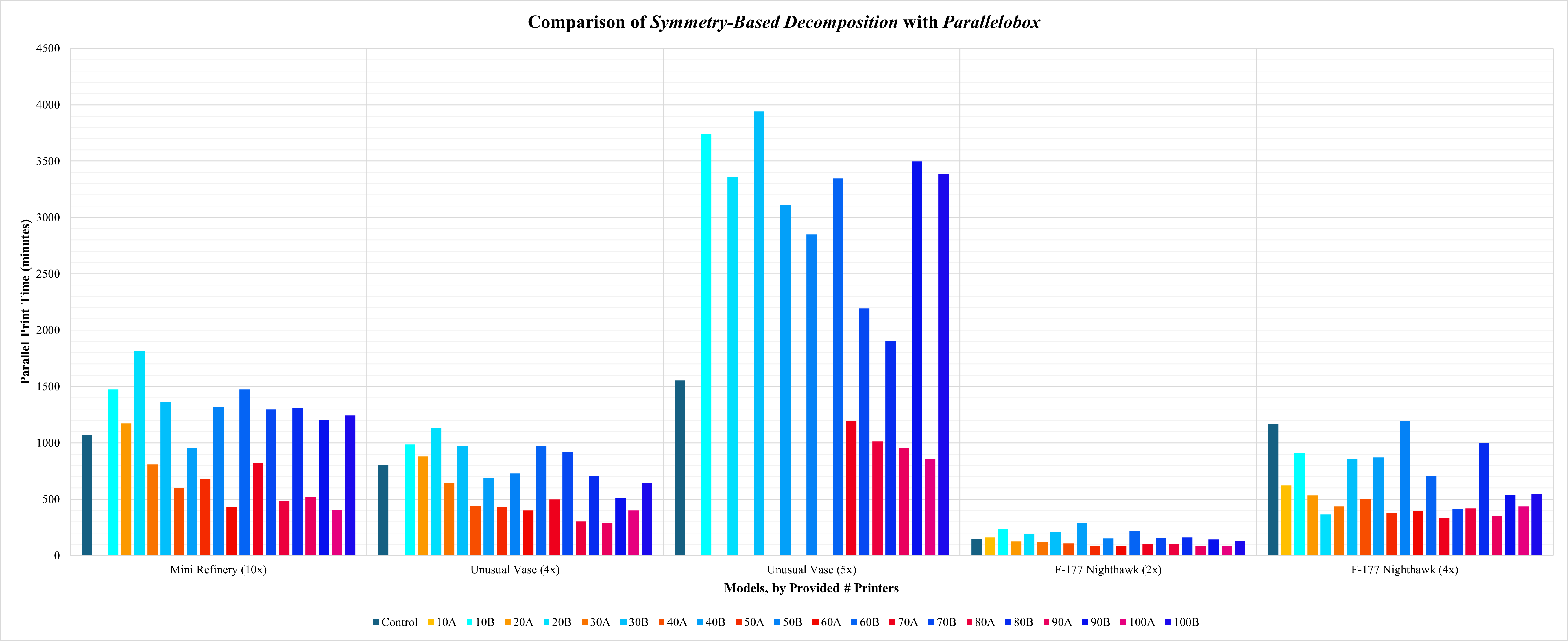}
    \includegraphics[width=\textwidth]{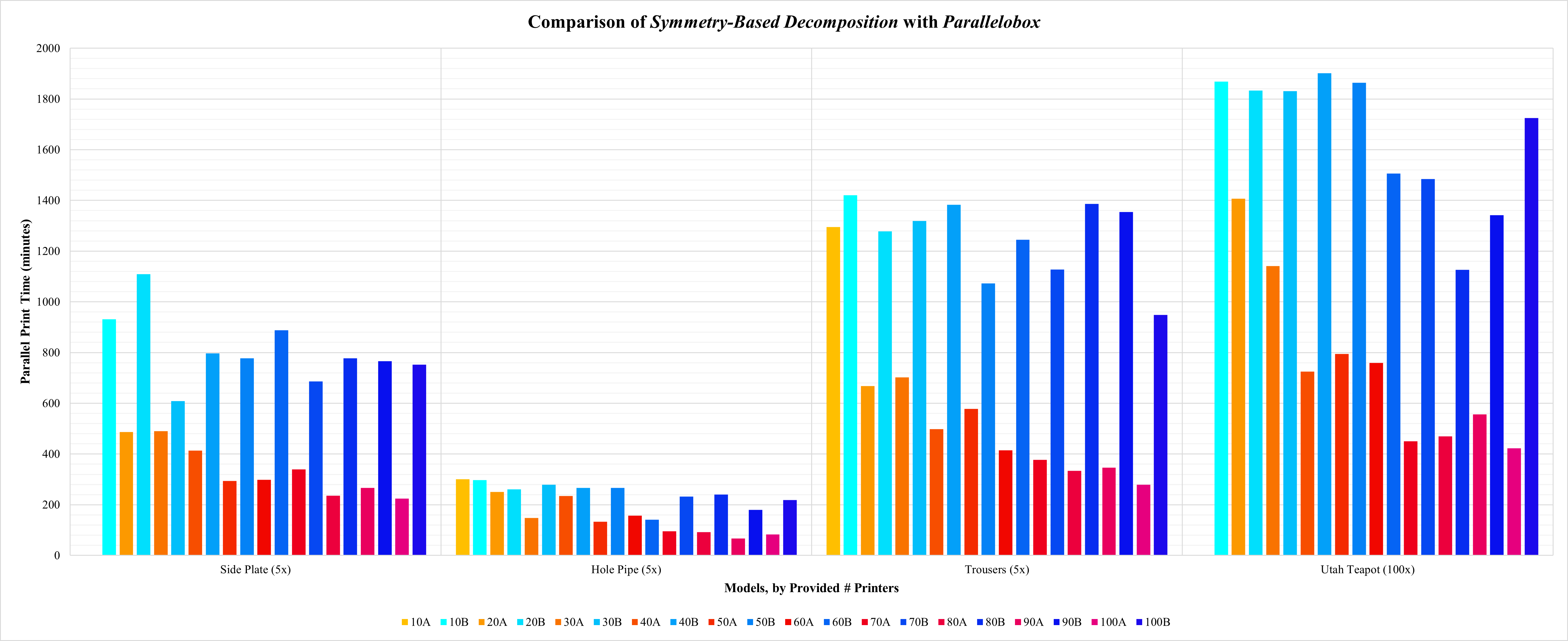}
    \includegraphics[width=\textwidth]{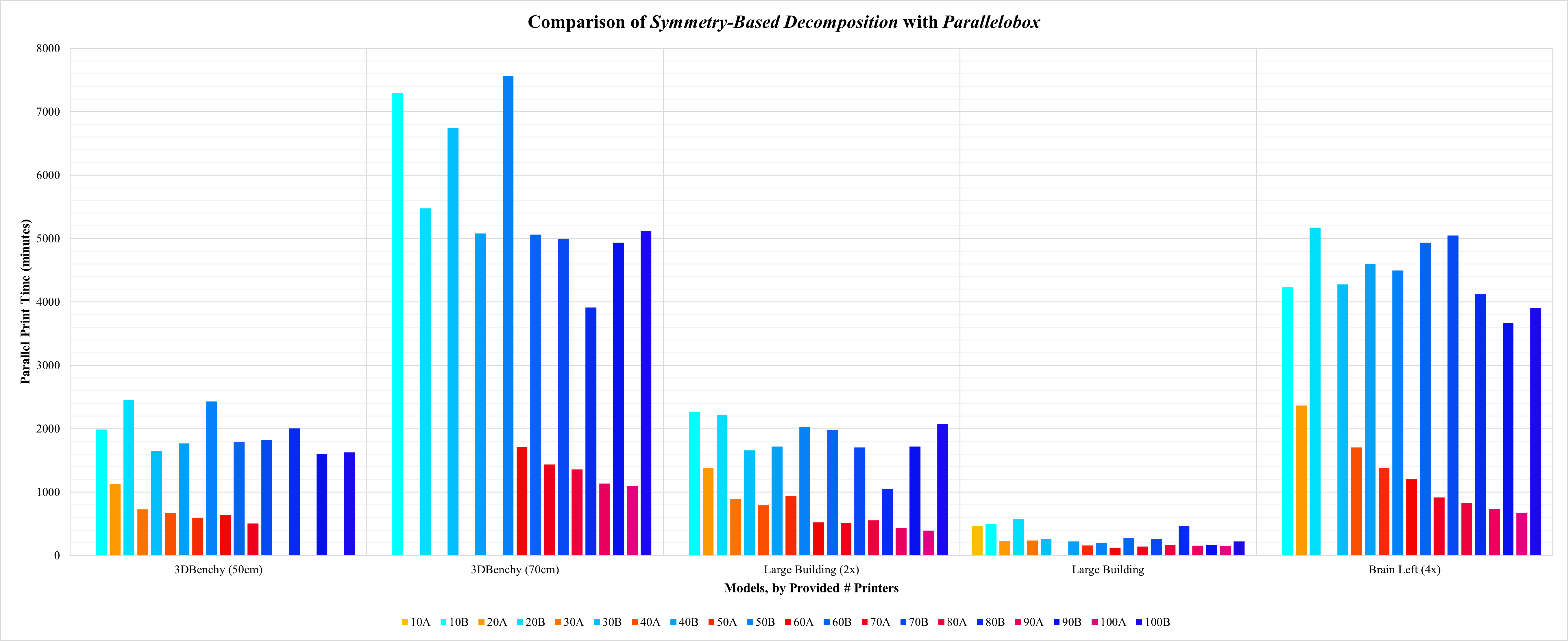}
\end{figure*}

\clearpage

\subsection{Appendix C: Parallelobox vs Symmetry-Based Decomposition Complete Data} \label{appendixC}

\begin{figure*}[hb!]
    \includegraphics[width=\textwidth]{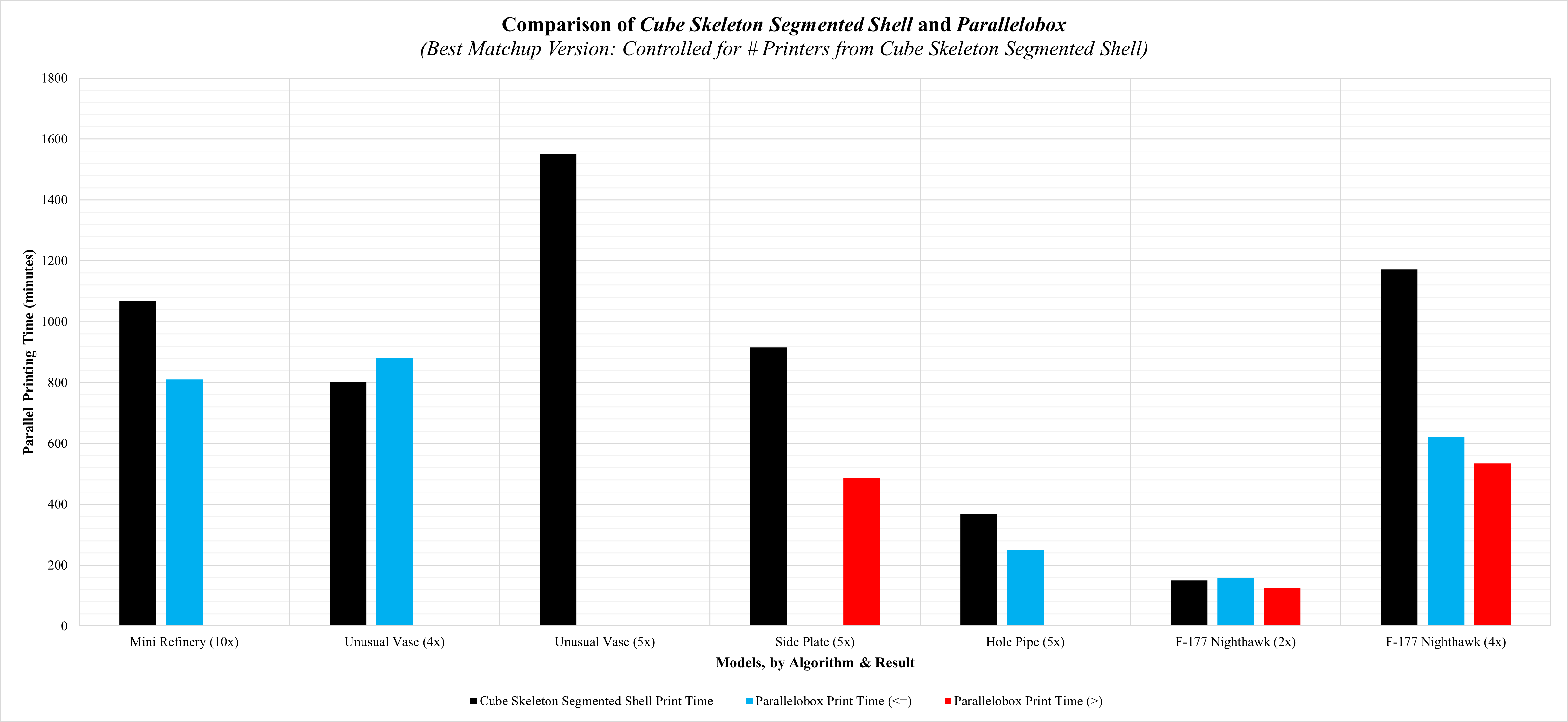}
    \includegraphics[width=\textwidth]{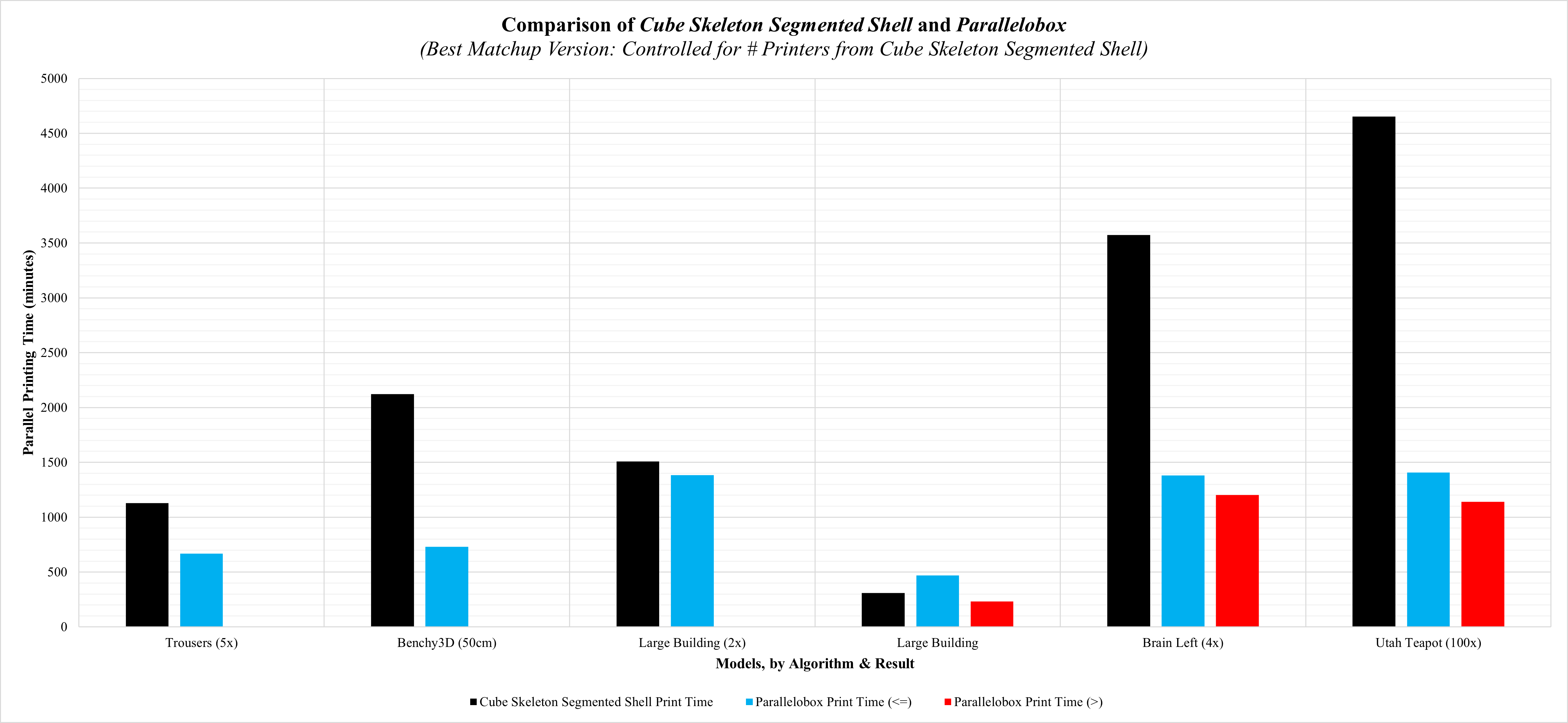}
\end{figure*}
%




\vfill


\end{document}